\documentclass{emulateapj}
\usepackage{graphicx}
\usepackage{epstopdf}
\usepackage{color}
\begin{document}
\submitted{Accepted by ApJ 2009 October 7}
\shorttitle{The Benchmark Companion HD 114762B}
\title{The Benchmark Ultracool Subdwarf HD 114762B: \\ A Test of Low-Metallicity Atmospheric and Evolutionary Models$^1$}
\author{Brendan P.\ Bowler,\altaffilmark{2} Michael C. Liu,\altaffilmark{3} and Michael C. Cushing}
\affil{Institute for Astronomy, University of Hawai`i \\ 2680 Woodlawn Drive, Honolulu, HI 96822, USA}
\email{bpbowler@ifa.hawaii.edu}

\altaffiltext{1}{Some of the data presented herein were obtained at the W.M. Keck Observatory, which is operated as a scientific partnership among the California Institute of Technology, the University of California and the National Aeronautics and Space Administration. The Observatory was made possible by the generous financial support of the W.M. Keck Foundation.}
\altaffiltext{2}{Visiting Astronomer at the Infrared Telescope Facility, which is operated by the University of Hawaii under Cooperative Agreement no. NCC 5-538 with the National Aeronautics and Space Administration, Science Mission Directorate, Planetary Astronomy Program.}
\altaffiltext{3}{Alfred P. Sloan Research Fellow}

\begin{abstract}

We present a near-infrared spectroscopic study of HD 114762B, the latest-type metal-poor companion discovered to date and the only ultracool subdwarf with a known metallicity, inferred from the primary star to be [Fe/H] = --0.7.  We obtained a medium-resolution (R $\sim$ 3800) Keck/OSIRIS 1.18-1.40 $\mu$m spectrum and a low-resolution (R $\sim$ 150) IRTF/SpeX 0.8-2.4 $\mu$m spectrum of HD 114762B to test atmospheric and evolutionary models for the first time in this mass-metallicity regime.  HD 114762B exhibits spectral features common to both late-type dwarfs and subdwarfs, and we assign it a spectral type of d/sdM9 $\pm$ 1.  We use a Monte Carlo technique to fit PHOENIX/$GAIA$ synthetic spectra to the observations, accounting for the coarsely-gridded nature of the models.  Fits to the entire OSIRIS $J$-band and to the metal-sensitive $J$-band atomic absorption features (\ion{Fe}{1}, \ion{K}{1}, and \ion{Al}{1} lines) yield model parameters that are most consistent with the metallicity of the primary star and the high surface gravity expected of old late-type objects.  The effective temperatures and radii inferred from the model atmosphere fitting broadly agree with those predicted by the evolutionary models of Chabrier \& Baraffe, and the model color-absolute magnitude relations accurately predict the metallicity of HD 114762B.  We conclude that current low-mass, mildly metal-poor atmospheric and evolutionary models are mutually consistent for spectral fits to medium-resolution $J$-band spectra of HD 114762B, but are inconsistent for fits to low-resolution near-infrared spectra of mild subdwarfs.   Finally, we develop a technique for estimating distances to ultracool subdwarfs based on a single near-infrared spectrum.  We show that this ``spectroscopic parallax'' method enables distance estimates accurate to  $\lesssim$ 10\% of parallactic distances for ultracool subdwarfs near the hydrogen burning minimum mass.

\end{abstract}

\section{Introduction}

A complete understanding of low-mass stars and brown dwarfs is an important goal of astrophysics.  Low-mass stars (0.08 $M_{\odot}$ $\lesssim$ $M$ $\lesssim$ 0.5 $M_{\odot}$)  are the most numerous objects in our galaxy (\citealt{Lada:2006p14060}) and make up nearly half of its total mass.  
 Brown dwarfs are objects that form like stars but have insufficient mass to sustain core nuclear reactions ($M$ $\lesssim$ 0.08 $M_{\odot}$) and represent the link between low-mass stars and extra-solar giant planets.  Objects with spectral types of M7 and later are known as ``ultracool'' objects and encompass the lowest-mass stars and brown dwarfs.  Our understanding of these objects has dramatically improved over the last decade, but there remain several fundamental areas of parameter space that are relatively unexplored, one of which is metallicity.

Sub-solar metallicity low-mass stars and brown dwarfs are referred to as ultracool subdwarfs.  Like their solar-metallicity counterparts, ultracool subdwarfs have spectral energy distributions that peak in the near-infrared and are dominated by overlapping molecular absorption bands.  Their lower metallicities, however, dramatically alter their spectral properties compared to solar-metallicity objects and can ultimately provide insight into the influence of metallicity on the chemistry, condensate cloud formation, and temperature/pressure profiles in the atmospheres of ultracool objects.  Prominent spectral changes resulting from a reduced metallicity include suppressed $H$- and $K$-band fluxes (i.e. bluer near-infrared colors) due to increased collision-induced absorption by H$_2$ (CIA H$_2$; \citealt{Linsky:1969p3947}; \citealt{Borysow:1997p3945}), a suppression of metal-oxide bands, and an enhancement of metal hydride bands (\citealt{Bessell:1982p13345}).  Ultracool subdwarfs exhibit large space motions consistent with thick disk or halo kinematics (\citealt{Burgasser:2007p575}), indicating that these objects are probably quite old (see, e.g., \citealt{Helmi:2008p17636}).

\begin{deluxetable}{lcc}
\tabletypesize{\scriptsize}
\tablewidth{0pt}
\tablecolumns{3}
\tablecaption{Age and Metallicity of HD 114762A: \\ Literature Search \label{metallicity} (1995 -- 2008)}
\tablehead{
\colhead{Reference}    &  \colhead{ Age (Gyr) }   &  \colhead{ [Fe/H] }   
}

\startdata

\citet{Haywood:2008p11758}  & 12.4 (2.8)   &  \nodata     \\
\citet{Holmberg:2007p11970} & 10.6 (2.2)  &  --0.76  \\
\citet{Gonzalez:2007p11922}   & \nodata   & --0.657 (0.051) \\
\citet{Zhang:2006p197}    & 14.1 &      --0.80 (0.10)  \\
\citet{Reddy:2006p11923}  &  11.3 (3.0)   &  --0.71  \\
\citet{Saffe:2005p183}   &  11.8 (3.6)   &   \nodata  \\
\citet{Valenti:2005p11833}   &   7.7 (2.3)   &  --0.65 (0.02)  \\
\citet{Santos:2004p5955}      & \nodata  &  --0.70 (0.04)    \\
\citet{Gratton:2003p6816}   &  \nodata & --0.79$_\mathrm{Fe \ I} $ \\
\citet{Laws:2003p12112} & 14 (2)  &  \nodata \\
\cite{Heiter:2003p11754}  &  \nodata  &  --0.78 (0.06) \\
\cite{Sadakane:2002p12239} & \nodata & --0.74 (0.07) \\
\citet{Fulbright:2000p11924}  &  \nodata  &  --0.7 (0.08) \\
\citet{Lachaume:1999p11806} & 11 (1)  & \nodata \\
\citet{Thevenin:1999p11819} & \nodata & --0.72$_\mathrm{LTE}$; --0.56$_\mathrm{NLTE}$ \\
\citet{Clementini:1999p11805}  &  \nodata  &  --0.66 (0.07) \\
\citet{Gonzalez:1998p5871}  &  14 (2) & --0.6 (0.06)    \\
\citet{Fuhrmann:1998p12487} &   \nodata   &  --0.71 ($\sim$0.05) \\
\citet{Ng:1998p5790}  &  10.8 (0.8)  & \nodata  \\
\citet{Henry:1997p5745}  &  5   &\nodata  \\
\cline{1-3} \\
\textbf{Adopted Values\tablenotemark{a}}  &  \textbf{11 (3)}  & \textbf{--0.71 (0.07)} \\

\enddata
\tablenotetext{a}{The adopted age and metallicity represent the mean and rms value of the listed quantities.}

\end{deluxetable}

The number of spectroscopically-confirmed ultracool subdwarfs has rapidly increased from the first identification over a decade ago (LHS 377, with a spectral type of sdM7: \citealt{Monet:1992p2839}; \citealt{Gizis:1997p80}) to the $\sim$ 45 currently known, the vast majority of which have been found in the past five years (see Table 7 of \citealt{Burgasser:2007p575}; \citealt{Burgasser:2008p2475}).  Many discoveries have been made through searches of large near-infrared and proper motion surveys including the Deep Near Infrared Survey of the Southern Sky (\citealt{Epchtein:1997p4581}), the Two Micron All Sky Survey (2MASS, \citealt{Skrutskie:2006p589}), and the Digitized Sky Survey (\citealt{Lepine:2005p70}).  For example, \citet{Lepine:2008p2934} recently identified  23 new M-type ultracool subdwarfs through template matching to Sloan Digital Sky Survey optical spectra.  Many more field ultracool subdwarfs are expected to be revealed with the next generation of deep, multi-epoch all-sky surveys such as the Panoramic Survey Telescope \& Rapid Response System (\citealt{Kaiser:2002p14879}) and the Large Synoptic Survey Telescope (\citealt{Tyson:2002p14903}).

The absolute metallicities of ultracool subdwarfs are currently unknown.  A commonly-used technique for estimating the metallicities of late-type objects is to fit synthetic spectra to their red-optical spectra and to adopt the metallicity of the best-fitting model (\citealt{Schweitzer:1999p563}; \citealt{Lepine:2004p559}; \citealt{Burgasser:2007p575}).   Evolutionary models have also been used to estimate the metallicities of ultracool subdwarfs in color-color  and color-magnitude diagrams (\citealt{Scholz:2004p580}; \citealt{Burgasser:2008p2475}; \citealt{Dahn:2008p14767}; \citealt{Schilbach:2009p14779}).  It is unclear how reliable these methods are, however, because models in this low-temperature, low-metallicity regime have not been tested.  There is thus a growing need for the discovery of ultracool subdwarfs with independently derived metallicities.

Benchmark objects whose fundamental parameters can be independently determined enable valuable tests of atmospheric and evolutionary models.  Several solar metallicity benchmark systems have already been used for such purposes.  One of the best examples is the HD 130948ABC system (\citealt{Potter:2002p5031}), which contains two L4 brown dwarfs in a hierarchical triple system with a G2V primary.  In this case the age and metallicity of the primary star along with the luminosities and dynamical masses of the brown dwarfs have been directly measured and used to test brown dwarf evolutionary models (\citealt{Dupuy:2009p15627}).  At sub-solar metallicities, however, there have not been any benchmark studies using very low-mass objects.  One possibility would be to use the latest-type objects in globular clusters, which comprise coeval stellar populations with uniform chemical compositions.  Their great distances, however, currently prevent spectroscopic studies of the lowest-mass members for which even photometric detections have proven difficult to obtain (\citealt{Richer:2002p14764}; \citealt{Richer:2006p14724}; \citealt{Richer:2008p14712}).  Another approach would be to use nearby ultracool subdwarfs in binary systems as sub-solar metallicity benchmarks.  Companions to field subdwarfs could also act as metallicity calibrators for the field ultracool subdwarf population.   This technique relies on the conservative assumption that the components of binary systems have the same metallicity.  Observations of binaries in the field (\citealt{Desidera:2004p109}) and in the Hyades (\citealt{Paulson:2003p3765}) have revealed that the components had differential [Fe/H] abundances of $\lesssim$ 0.02-0.04 dex, implying that the assumption of identical metallicity is indeed reasonable.   Unfortunately, most of the known ultracool subdwarfs are isolated objects and searches for low-mass secondaries to metal-poor stars have yielded surprisingly few bona fide companions (\citealt{ZapateroOsorio:2004p5007}; \citealt{Chaname:2004p4964}; \citealt{Riaz:2008p4770}).

HD 114762B is currently the latest-type object known that is gravitationally bound to a metal-poor star (sdF9; HD 114762A).   It was discovered and confirmed as a proper motion companion in a Lick Observatory adaptive optics search for stellar and substellar companions to exoplanet host stars (\citealt{Patience:2002p186}) and was later rediscovered in a similar study using CFHT AO (\citealt{Chauvin:2006p178}).  HD 114762B is located 3$\farcs$3 (128 AU at 38.7 pc; \citealt{vanLeeuwen:2007p12454}) from its primary, which also harbors a high-mass planet, brown dwarf, or low-mass star, depending on the system's inclination ($M_\mathrm{P}$sin$i$ = 11.68 $\pm$ 0.96 $M_\mathrm{Jup}$; \citealt{Latham:1989p5199}; \citealt{Cochran:1991p5246}; \citealt{Hale:1995p6074}; \citealt{Butler:2006p3743}).  The presence of a massive exoplanet candidate has led to numerous studies of the primary's metallicity ([Fe/H]) and age, which have been well constrained to be --0.71 $\pm$ 0.07 dex and 11 $\pm$ 3 Gyr, respectively\footnote{The quoted errors are rms values based on a literature search rather than formal uncertainties.}  (Table \ref{metallicity}).  Age estimates in the literature for HD 114762A are based on theoretical isochrones and chromospheric activity levels.  This system may also harbor a debris disk based on optical polarimetry and ISO far-IR photometry (\citealt{Saffe:2004p187}). 

A Keck adaptive optics/NIRSPEC $J$-band spectrum of HD114762B was presented by \citet{Patience:2002p186}.  They found inconsistent spectral features that appeared similar to a late-M/early L spectrum based on the strength of the 1.4 $\mu$m steam band, but also resembled a mid-M spectrum based on the strengths of the 1.189 $\mu$m \ion{Fe}{1} and 1.313 $\mu$m \ion{Al}{1} neutral metal lines and on a comparison to $J$-band spectra of other M dwarfs.  However, the dearth of ultracool subdwarfs known at the time of discovery prevented a comparative analysis with other metal-poor objects, and the nature of their observations made the analysis difficult because slit-based spectroscopy with adaptive optics does not preserve the continuum shape  (\citealt{Goto:2002p495}; \citealt{Goto:2003p13421}; \citealt{McElwain:2007p13510}).

Here we present a near-infrared spectroscopic study of HD 114762B.  Our goals are threefold: ($1$) to better characterize this object through a comparison to known subdwarfs, ($2$) to test atmospheric models by comparing the results of the best-fitting models to the known metallicity of this object, and (3) to study the consistency of the predictions from atmospheric and evolutionary models.  For our tests we use the $GAIA$ grid of synthetic spectra (\citealt{Brott:2005p301}, based on the PHOENIX atmospheric model code) and the \citet[CB97]{Chabrier:1997p2767}, \citet[BCAH97]{Baraffe:1997p582}, and \citet[BCAH98]{Baraffe:1998p160}  low-metallicity evolutionary models (all based on the same input physics).  These models are widely used throughout the literature and are therefore important to validate.

In $\S$2 we present our observations and the methods for extracting the spectra from the reduced data.  In $\S$3 we describe the results of the $\chi^2$ fitting of the models to the data and the way in which we derive the fundamental parameters from the atmospheric and evolutionary models.  Finally, we summarize our results and discuss the implications of our work in $\S$4.

\section{Observations and Data Reduction}

\subsection{OSIRIS Spectrum}

We obtained medium-resolution (R $\sim$ 3800; R $\equiv$ $\lambda/\Delta\lambda$) $J$-band spectra of HD 114762B with the OH-Suppressing InfraRed Integral-field Spectrograph (OSIRIS; \citealt{Larkin:2006p5567}) using the natural guide star adaptive optics system at Keck II on the night of 2008 Jan 14 UT.  The weather was poor with patchy cirrus clouds.  Using HD 114762A as the natural guide star, we obtained four exposures of HD 114762B with an integration time of 600 s and one with an integration time of 120 s for a total integration time of 42 min.  Immediately afterwards we observed the A0V star HD 116960 for telluric correction with four exposures of 60 s each.  We used the 50 mas pixel scale for these observations.  OSIRIS uses a lenslet array to sample a rectangular region of sky, producing 16 $\times$ 64 individual spectra which are later merged into a data cube covering both spatial and dispersion directions.  The data were reduced and wavelength calibrated using the OSIRIS data pipeline (\citealt{Krabbe:2004p13521}).  We extracted individual spectra from each observation by performing a robust third-order polynomial fit to the centroids of each target in the dispersion direction for both the the x- and y-coordinates.  We then performed aperture photometry centered on the polynomial fit positions at each wavelength slice using an aperture radius of 3 pixels (0$\farcs$15) and an inner and outer sky annulus of 7 pix (0$\farcs$35) and 10 pix (0$\farcs$50), respectively.  The typical FWHM was 65 mas.

Individual spectra were combined by first multiplicatively scaling them to the median-combined spectrum and then computing the median of the scaled spectra.  Measurement errors were determined by computing standard errors of the median of the scaled spectra.  The final spectrum was corrected for telluric absorption using a general version of the Spextool reduction package (\citealt{Vacca:2003p497}; \citealt{Cushing:2004p501}), which was originally written for the near-infrared spectrograph SpeX at IRTF.  

\begin{figure*}
  \plotone{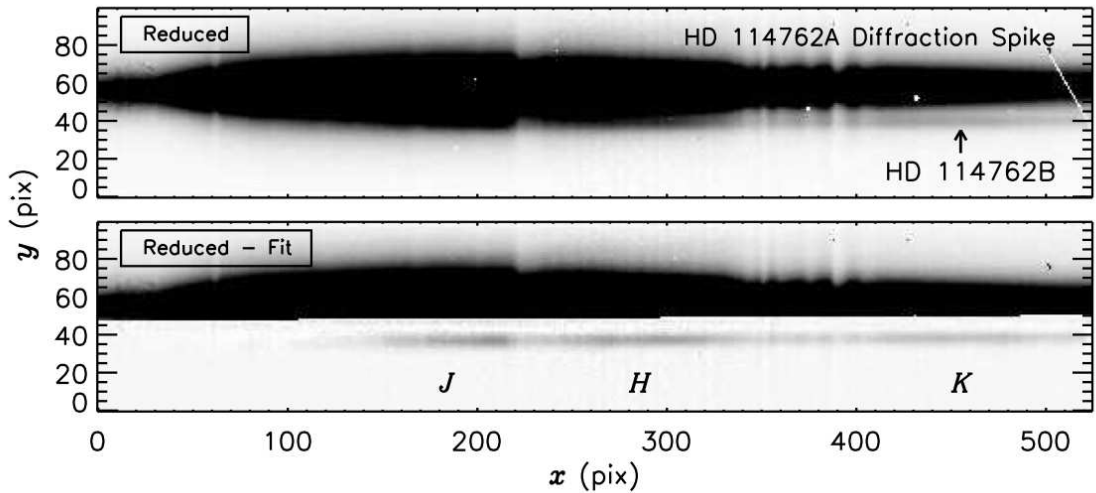}
  \caption{Extraction of the HD 114762B spectrum from the SpeX/prism low-resolution data.  The upper panel shows a single sky-subtracted and flat-fielded two-dimensional spectrum of the companion HD 114762B and the very bright diffraction spike of the primary star HD 114762A.  The $x$-axis is the dispersion direction and the $y$-axis is the spatial direction (the spatial scale is 0$\farcs$15 pix$^{-1}$).  The lower panel shows the image after fitting and subtracting the diffraction spike.  The approximate locations of the $J$, $H$, and $K$ bands are labeled.   \label{f1} } 
\end{figure*}

\subsection{SpeX Prism Spectrum}

\subsubsection{Observations}

We obtained low-resolution (R $\sim$ 150) 0.8-2.4 $\mu$m near-infrared spectra of HD 114762B using the near-infrared SpeX spectrograph (Rayner et al. 2003) in prism mode at IRTF on 2008 Jan 28 UT.  Conditions were not photometric but the seeing was good with a typical FWHM of ~0$\farcs$8 throughout the observations over an airmass range of 1.013 to 1.001.  We chose a slit length of 15$\arcsec$ and a slit width of 0$\farcs$5.  The close angular separation (3$\farcs$3) and the high contrast ($\Delta$$J$ $\sim$ 7.6, \citealt{Patience:2002p186}) between the primary and the companion made the observations difficult.  With a slit orientation in the radial direction from the primary we would have needed to position the target near one end of the slit in order to avoid contamination from the primary, thereby losing a fraction of the flux from the companion.  We therefore chose a slit orientation roughly orthogonal to the line connecting the primary to the companion in order to capture the majority of the flux of HD 114762B near the center of the slit.  Unfortunately a diffraction spike from the primary was situated $\sim$ 3$\arcsec$ from the companion so our observations include a spectrum of a bright diffraction spike along with that of the much fainter science target (Figure \ref{f1}, top panel).  

The data were taken by nodding on and off the target in the $ABBA$ configuration by about 13$\arcsec$, and consist of 11 cycles of 120 s each for a total integration time of 22 min.  We subtracted the raw data in pairs to remove the night sky lines, dark current, and  bias offset.  The subtracted pairs were then divided by a flat-field which was created by first scaling to a median-combined flat and then taking the median of the scaled frames.  A slight residual sky signal was apparent across the entire spectrum.  We removed the residual sky at each column by fitting a line to the median-combined flux near both ends of the slit in the spatial direction and then subtracting off the fitted line.  This procedure simply flattened the data to zero at both ends of the spatial profile.  The effect of removing the residual sky was small and the extraction of the companion spectrum (see below) was done with this flattening and without it; the result on the final spectrum was negligible.  We accounted for chip defects in the SpeX infrared array detector by using the bad pixel mask provided in Spextool.

\subsubsection{Extracting the Spectrum}

Extracting the one-dimensional spectrum from the two-dimensional reduced data was not straightforward.  The main difficulty lay in determining the spatial shape of the diffraction spike spectrum whose tail the companion was immersed.  The peak of the diffraction spike was $\sim$ 3$\arcsec$ from that of the companion in the spatial direction.  Ideally we would like to know the exact functional form of the bright tail so that we could simply subtract it off at each column.  Practically this meant fitting simple functions to the tail to determine the one that best approximated the data.  

The spatial profile of the diffraction spike was slightly asymmetric so we could not use the opposite tail as a model.  We first tested a method of spline interpolation across the location of the science target.  This technique was highly sensitive to both the ``tension'' parameter of the spline as well as the choice of anchor points along the tail; the result was a noisy and over-subtracted companion spectrum.  We tried constant values for the tension parameter as well as tension values estimated from the best-fitting spline interpolation over the same region on the opposite tail.  Both methods produced poor results.   We tested a wide variety of standard functions including a Gaussian, a Moffat, and a number of polynomials (from 2$^\mathrm{nd}$ to 6$^\mathrm{th}$ order) in both normal- and log-space.  The quality of each fit was judged based on a visual inspection of the fit at each column and of the resulting subtracted two-dimensional spectrum.  The best result was a third-order polynomial fit to the logarithm of the intensity values.  The functional form in linear flux units is 

\begin{equation}
I(y) = 10^{a+by+cy^2+dy^3}, 
\end{equation}

\noindent where $I$ is the intensity at each pixel, $y$ is the pixel coordinate in the spatial direction, and the values of $a$, $b$, $c$, and $d$ are the constants that are fit for at each column.   In each fit the data were median-combined over 11 pixels in the dispersion direction to create a higher signal-to-noise spatial profile.  Anchor points used in the fit were centered around the peak of the companion and were adjusted for a small ($\sim$ 3 pixel) shift caused by a slight rotation of the data on the array.  The adjustment was made by fitting Gaussians to the diffraction spike spectrum at each column, then fitting a line to the peak of the Gaussians in the dispersion direction, and then finally offsetting the anchor points from the linear fit.  This ensured that the anchor points did not move with respect to the spectrum of the science target.  The fit was then scaled to the values of the original (non-median-combined) data at the central column by a scale factor that accounted for small differences in height of the spatial profile.   The scaling was performed using \texttt{MPFITEXPR}, a robust non-linear least squares curve fitting routine in IDL (\citealt{Markwardt:2009p14854}).  Finally, the scaled fit was subtracted from the tail of the diffraction spike thereby revealing the spatial profile of HD 114762B (Figure \ref{f1}, bottom panel).  This procedure was applied to every column across the entire spectrum for each reduced frame.

One-dimensional spectra were obtained for each individual frame by summing the flux within a width of 5 pixels (0$\farcs$75) from either side of the peak of the spatial profile of the target spectrum at each column.  To ensure a smooth aperture radius along the dispersion direction, the peak of the spatial profile was determined by first fitting a Gaussian to the spatial profile at each column and then fitting a line in the dispersion direction to the peak of each Gaussian fit.  To combine the spectra from the individual frames the data were scaled to the median-combined spectrum over the $H$-band and were median-combined again.  Telluric correction was performed using the Spextool IRTF data reduction and analysis package (v.3.4, \citealt{Vacca:2003p497}; \citealt{Cushing:2004p501}).  We performed several trial extractions to test the influence of the aperture radius  and anchor points of the fitted function on the final coadded spectrum.  The aperture radii were tested with values of 1, 3, 5, 7, and 10 pixels.  The shape of the spectra changed very little, especially for aperture radii $>$ 3 pix.  The effect of the anchor point positions were also tested by examining the extracted spectra and the subtracted two-dimensional spectral images of the companion.  Large changes of $\sim$ 5 pixels in the anchor point positions strongly influenced the overall shape of the resulting spectra and introduced considerable noise.  We chose anchor points that produced the best  fits based on a visual inspection of the subtracted images and on a comparison of extracted spectra using different anchor point regions.

\begin{figure}
  \resizebox{3.45in}{!}{\includegraphics{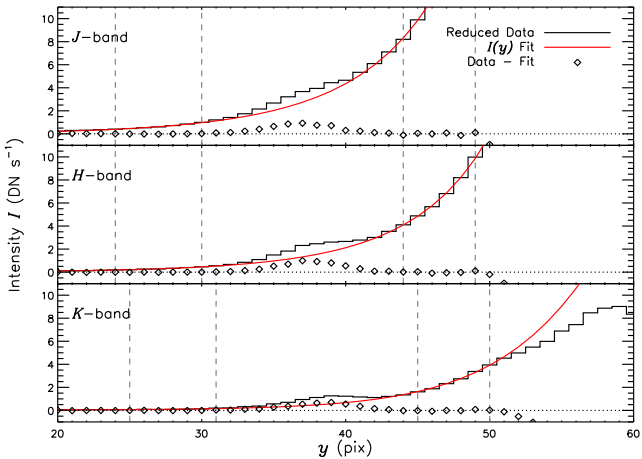}}
  \caption{Extraction of the HD 114762B spectrum from the SpeX/prism low-resolution data.  The spatial scale of the infrared array is 0$\farcs$15 pix$^{-1}$.  The science target is the small bump seen on the steeply rising flux from the primary star's diffraction spike.  At each column we fit a function (red line, Equation 1) to the spatial profile of the diffraction spike using anchor points (demarcated by gray dashed lines) that were allowed to shift with the peak of the diffraction spectrum across the array.  The fit was subtracted from the bright tail, revealing the spectrum of the companion (plotted as diamonds).   The contrast between the science target and the diffraction spike was worse in the $J$ band than in the $K$ band and may have led to a slight oversubtraction of flux at short wavelengths, especially below $\sim$ 1 $\mu$m.  See $\S$ 2.2.2 for details of the extraction process and the functional form of the fit.   \label{f2} \\ \\ } 
\end{figure}

The contrast between the companion and the diffraction spike spectra was especially large shortward of $\sim$ 1 $\mu$m (Figure \ref{f2}).  In this region small changes in the goodness-of-fit of Equation 1 corresponded to large variations in the resulting subtracted profile.  For that reason,  \emph{the companion spectrum is probably not reliable for $\lambda$ $\lesssim$ 1 $\mu$m.}  While general absorption features are apparent, the flux levels are probably not trustworthy.

\subsection{Flux Calibration}

We flux-calibrated the OSIRIS and SpeX spectra using published photometry.  Differential $J$, $H$, and $K_\mathrm{S}$ magnitudes between HD 114762A and HD 114762B are listed in Table 4 of \citet{Patience:2002p186}.  Their imaging observations were obtained using Lick Observatory's IRCAL camera with adaptive optics.  We used the differential magnitudes combined with 2MASS photometry of the primary to compute the IRCAL magnitudes of HD~114762B:

\begin{equation}
J_\mathrm{com}^\mathrm{IRCAL} = \Delta J^\mathrm{IRCAL}  + (J_\mathrm{pri}^\mathrm{2MASS} + J_\mathrm{corr}),
\end{equation}

\noindent where $J_\mathrm{corr}$ is the correction term used to transform the 2MASS $J$-band magnitude to the IRCAL photometric system.  The same procedure was applied to the $H$ and $K_\mathrm{S}$ bands.  The correction term was derived by calculating the differential magnitude of an F9 star in both photometric systems.  The choice of an F9 star is based on the spectral type of HD 114762A (sdF9), which allows us to avoid a color correction between the filter systems. We performed synthetic photometry on the infrared spectrum of HD 27383 (F9V) which we obtained from the IRTF spectral library (Rayner et al., submitted).   The correction term for the $J$ band is

\begin{eqnarray}
J_\mathrm{corr} = -2.5 \log\left(\frac{ \int \lambda   f_{\lambda}^\mathrm{F9} T_{J}^\mathrm{IRCAL}(\lambda)  d\lambda }{\int \lambda  f_{\lambda}^\mathrm{Vega}  T_{J}^\mathrm{IRCAL}(\lambda) d\lambda}\right) + \nonumber \\
2.5\log\left(\frac{ \int \lambda  f_{\lambda}^\mathrm{F9}  T_J^\mathrm{2MASS}(\lambda) d\lambda }{\int \lambda  f_{\lambda}^\mathrm{Vega}  T_J^\mathrm{2MASS}(\lambda) d\lambda}\right).
\end{eqnarray}

\noindent Here $T_J^\mathrm{IRCAL}$ and $T_J^\mathrm{2MASS}$ are the transmission curves of the $J$ filter in the IRCAL and 2MASS systems, respectively, while $f_{\lambda}^\mathrm{F9}$ and  $f_{\lambda}^\mathrm{Vega}$ are the flux densities of the F9 star and Vega.  The 2MASS transmission filters include the filter profile together with the atmospheric transmission profile, so we multiplied the IRCAL filter transmissions with an infrared telluric profile generated using ATRAN (\citealt{Lord:1992p5602}).  For Vega fluxes throughout this work we used the flux-calibrated spectrum of Vega provided in Spextool.   The correction terms to convert from the 2MASS to the IRCAL systems were $<$ 0.01 mags for all three bands, which is smaller than the errors in the 2MASS photometry of HD 114762A and much smaller than the errors in the IRCAL differential magnitudes.  The corrections are therefore negligible and we consider the 2MASS and IRCAL filters to be the same systems in this work.  The photometry for HD 114762B is listed in Table \ref{synthphot}.   Errors are derived from the published magnitude difference errors and the 2MASS photometric errors, added in quadrature.

We flux calibrated the SpeX data by computing the multiplicative constant that scaled the spectrum to the flux density level that matched the $K_\mathrm{S}$-band magnitude listed in Table \ref{synthphot}.  The $K_\mathrm{S}$ band was chosen because any systematic errors caused by the extraction of the spectrum would be minimized in this bandpass (see $\S$ 2.2.2).  The flux calibration scaling factor $C_\mathrm{fc}$ was computed in a Monte Carlo fashion.  An artificial spectrum was generated by adding to the data noise drawn from a Gaussian distribution with a  standard deviation equal to the measurement error at each pixel.  The flux calibration scaling factor $C_\mathrm{fc}$ was then computed for each artificial spectrum in the following manner:

\begin{equation}
C_{\mathrm{fc}, i} = \frac{ \int \lambda  f_{\lambda}^\mathrm{Vega}  T_{K_\mathrm{S}}(\lambda) d\lambda }{ \int \lambda  f_{\lambda, i}^\mathrm{MC}  T_{K_\mathrm{S}}(\lambda) d\lambda } \times 10^{-0.4K_{\mathrm{S},  i}},
\end{equation}

\noindent where  $f_{\lambda, i}^\mathrm{MC}$ is the Monte Carlo-generated spectrum for trial $i$, and $f_{\lambda}^\mathrm{Vega} $ and $T_{K_\mathrm{S}}(\lambda)$ are the same as in Equation 3.  For each trial $i$, a new $K_{\mathrm{S},  i}$ magnitude was drawn from a Gaussian distribution with a mean value equal to the $K_\mathrm{S}$-band magnitude from Table \ref{synthphot} and a standard deviation equal to the photometric uncertainty.  The mean flux calibration scaling factor $\overline{C}_\mathrm{fc}$ and its error $\sigma_{\overline{C}_\mathrm{fc}}$ were obtained from the distribution of $C_\mathrm{fc}$ values for 10$^4$ trials.  The resulting values of $\overline{C}_\mathrm{fc}$ and $\sigma_{\overline{C}_\mathrm{fc}}$ were 1.10 and 0.10, respectively.  The error in the final flux calibration level is roughly 9\% and is dominated by the $K_\mathrm{S}$-band photometric uncertainty.

The OSIRIS spectrum does not cover the entire 2MASS $J$ band so we used the SpeX spectrum for flux calibration.  We smoothed the OSIRIS spectrum to the resolution of the SpeX data and then scaled it to the same flux level.  We assume an identical flux calibration uncertainty of 9\% for the OSIRIS spectrum based on the SpeX spectrum.

\begin{deluxetable*}{cccccccc}
\tabletypesize{\scriptsize}
\tablewidth{0pt}
\tablecolumns{8}
\tablecaption{HD 114762B Infrared Photometry (2MASS System) \label{synthphot}}
\tablehead{
\colhead{Type} & \colhead{$J$ ($\sigma_{J}$)}  &\colhead{$H$ ($\sigma_{H}$)}  &\colhead{$K_\mathrm{S}$ ($\sigma_{K_\mathrm{S}}$)}  &   \colhead{$J - H$ ($\sigma_{J-H}$)}         &   \colhead{$H - K_\mathrm{S}$ ($\sigma_{H - K_\mathrm{S}}$)}   &    \colhead{$J - K_\mathrm{S}$ ($\sigma_{J-K_\mathrm{S}}$)}       & \colhead{Ref}    
}
\startdata
Aperture Phot. & 13.74 (0.10) & 13.39 (0.10) & 13.01 (0.10) &     0.35 (0.14)   &  0.38 (0.14)  &  0.73 (0.15)  &  1, 2  \\
Synthetic Phot. & 13.99 (0.10)  &  13.45 (0.10) & 12.99 (0.10) &   0.54 ($<$ 0.01)    &   0.46 ($<$ 0.01)   &  1.00 ($<$ 0.01)   &  3       \\
\enddata
\tablerefs{(1) \citet{Patience:2002p186}; (2) \citet{Cutri:2003p5612}; (3) This work.}
\end{deluxetable*}

\begin{figure*}
  \plotone{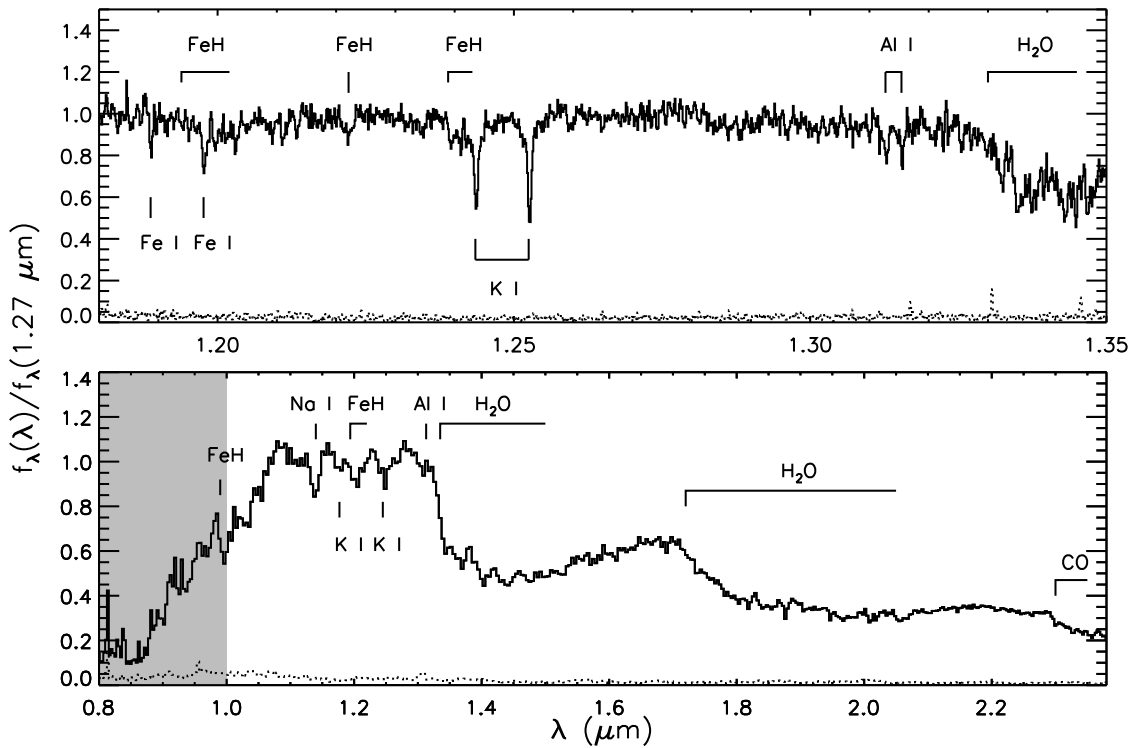}
  \caption{Line identification in HD 114762B from the medium-resolution OSIRIS spectrum (top) and the low-resolution SpeX/prism spectrum (bottom).  Major molecular bands are from H$_2$O, FeH, and CO, while prominent atomic features originate from \ion{K}{1}, \ion{Al}{1}, \ion{Na}{1}, and \ion{Fe}{1}.  Measurement errors for both spectra are shown as dotted lines.  The portion of the SpeX spectrum that may have been altered by the spectral extraction technique is demarcated by the gray shaded region (see \S 2.2.2).     \label{f3} \\ } 
\end{figure*}

\subsection{Synthetic Photometry}

We computed synthetic magnitudes and colors from our flux-calibrated SpeX spectrum and compared them to the published photometry to test whether the extraction technique altered the slope of the final spectrum.  Synthetic magnitudes were derived in a Monte Carlo fashion incorporating measurement errors and flux-calibration errors.  For each trial, an artificial spectrum was generated in the manner described in \S 2.3 and was multiplied by a flux-calibration scaling factor which was drawn from a Gaussian distribution with a mean value of $\overline{C}_\mathrm{fc}$ and a standard deviation of $\sigma_{\overline{C}_\mathrm{fc}}$.  Synthetic $J$, $H$, and $K_\mathrm{S}$ magnitudes were then computed using 2MASS transmission profiles.  This process was repeated 10$^4$ times.  The mean synthetic magnitudes and the standard deviations of the resulting distributions are listed in Table \ref{synthphot}.  Synthetic colors and errors were computed in a similar manner and are presented in the same table.

The synthetic colors and magnitudes from our SpeX spectrum disagree with those derived from the \citet{Patience:2002p186} differential photometry.  The disagreement appears to progressively worsen at shorter wavelengths in the form of a shallower spectral slope in the SpeX spectrum.  It is unclear whether this difference originates in the spectral extraction method used in this work or whether it originates with the published photometry.  A greater offset with decreasing wavelength, however, is consistent with an oversubtraction of flux in our extraction technique of the companion spectrum (\S 2.2.2).  The oversubtraction would be negligible for the low contrast $K$-band region but would be significant for the high contrast $J$-band region.  We accounted for this difference by dividing the SpeX spectrum into three sections and adjusting each section's relative flux to match the $J$-, $H$-, and $K_\mathrm{S}$-band photometry.  The sections were chosen to approximate the near-infrared bandpasses: a $J$ section (1.1-1.33 $\mu$m), an  $H$ section (1.5-1.8 $\mu$m), and a  $K$ section (2.0-2.35 $\mu$m).  We left the $K$-band spectrum stationary as it was probably least affected by the extraction process.  We refer to the resulting spectrum as the ``shifted SpeX spectrum" throughout this work.

\begin{figure*}
  \plotone{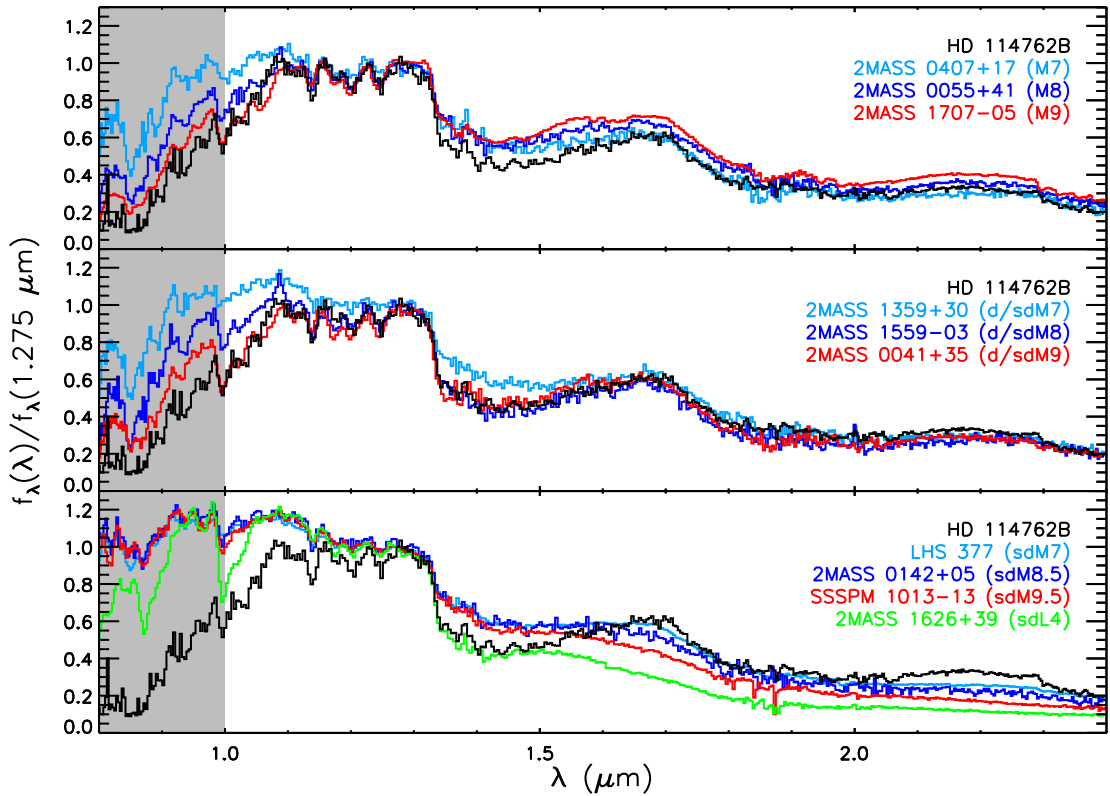}
  \caption{Comparison of HD 114762B (black) to spectra of late-type objects using the same instrument at IRTF (SpeX/prism).  The top panel is a comparison to late-type dwarfs, the middle to mild subdwarfs, and the bottom to subdwarfs.  The deep 1.4 and 1.9 $\mu$m H$_2$O steam bands are poorly matched by the dwarfs and the subdwarfs.  We assign a spectral type d/sdM9 $\pm$ 1 to HD 114762B based on its close resemblance to the mildly metal-poor object 2MASS 0041+35.  The data are from the SpeX Prism Spectral Library and the observations were originally presented by \citet{Burgasser:2004p574}, \citet{Burgasser:2004p564}, and \citet{McElwain:2006p5628}.  The full names of the objects used for comparison are, from top to bottom panels, 2MASS 04071296+1710474, 2MASS 00552554+4130184, 2MASS 17072343--0558249, 2MASS 13593574+3031039, 2MASS 15590462--0356280, 2MASS00412179+3547133, LHS 377, 2MASS 01423153+0523285, SSSPM 1013--1356, and 2MASS 16262034+3925190. The portion of the SpeX spectrum that may have been altered by the spectral extraction technique is demarcated by the gray shaded region (see \S 2.2.2).   \label{f4} \\ } 
\end{figure*}

\section{Results}

\subsection{Spectroscopic Characterization of HD 114762B}

The spectra of HD 114762B reveal that it is a late-type object with a spectral energy distribution dominated by strong overlapping molecular absorption bands, most notably the deep 1.4 and 1.9 $\mu$m H$_2$O steam bands.  The strongest atomic absorption lines and the major molecular absorption bands are labeled in Figure \ref{f3}.  The OSIRIS $J$-band spectrum shows a pseudo-continuum that originates mostly from H$_2$O but that also contains defining FeH band heads at 1.1939 and 1.2389 $\mu$m (\citealt{Phillips:1987p4323}; \citealt{Jones:1996p4544}; \citealt{Cushing:2003p272}; \citealt{McLean:2007p2553}) and a noticeable Q-branch feature at 1.222 $\mu$m (\citealt{Cushing:2003p272}).  Prominent atomic lines are from \ion{Fe}{1} at 1.1886/1.1887 (blended) and  1.1976 $\mu$m,  \ion{K}{1} at 1.2436 and 1.2526 $\mu$m,  and \ion{Al}{1} at 1.3127 and 1.3154 $\mu$m.  The SpeX spectrum exhibits a rich molecular band structure which includes the Wing-Ford FeH band at 0.99 $\mu$m (\citealt{Wing:1969p4145}), other strong FeH band heads at 1.239 and 1.194 $\mu$m, and the 2.3 $\mu$m CO band.  Neutral atomic lines are also visible, including a blended \ion{K}{1} doublet at 1.177 and 1.245 $\mu$m, a blended \ion{Al}{1} doublet at 1.314 $\mu$m, and a deep \ion{Na}{1} doublet at 1.14 $\mu$m.

In Figure \ref{f4} we compare the SpeX spectrum of HD 114762B to spectra of late-type dwarfs, to so-called ``mild subdwarfs'' which are believed to be only slightly metal-poor, and to subdwarfs in the top, middle, and bottom panels, respectively.  The spectra are from the SpeX Prism Spectral Library\footnote{Maintained by Adam Burgasser at http://www.browndwarfs.org/spexprism. } and are normalized at 1.275 $\mu$m.  HD 114762B closely resembles the M9 object 2MASS 17072343--0558249 (\citealt{McElwain:2006p5628}) between 1.00-1.35 $\mu$m, but has a deeper 1.4 $\mu$m H$_2$O steam band than any of the normal dwarfs.  A much better fit is obtained when compared to the mild subdwarfs.  The general shape of the spectral energy distribution matches quite well, except in the $K$ band, where the spectrum of HD 114762B appears slightly higher than the mild subdwarfs.  This may be caused by a slight oversubtraction at shorter wavelengths (see $\S$ 2.2.2) which, when normalized to 1.275 $\mu$m, would result in a raised $K$-band flux.  Finally, a comparison to subdwarf templates shows poor matches across the entire wavelength range.

\begin{figure}
  \resizebox{3.45in}{!}{\includegraphics{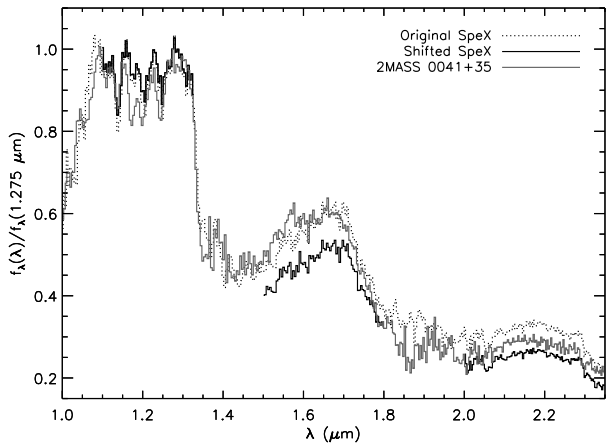}}
  \caption{SpeX spectrum shifted to the photometry of \citet[solid black]{Patience:2002p186} and compared to 2MASS 0041+35 (d/sdM9, gray), which is the best-fitting reference object to the original unshifted SpeX spectrum (dotted black).  All spectra are normalized to 1.275 $\mu$m.  The $H$-band flux of the shifted SpeX spectrum is significantly lower than that of 2MASS 0041+35, which is probably a result of scaling the spectral segments of HD 114762B to the published photometry which had large errors.\label{f5} \\} 
\end{figure}

\begin{figure}
  \resizebox{3.45in}{!}{\includegraphics{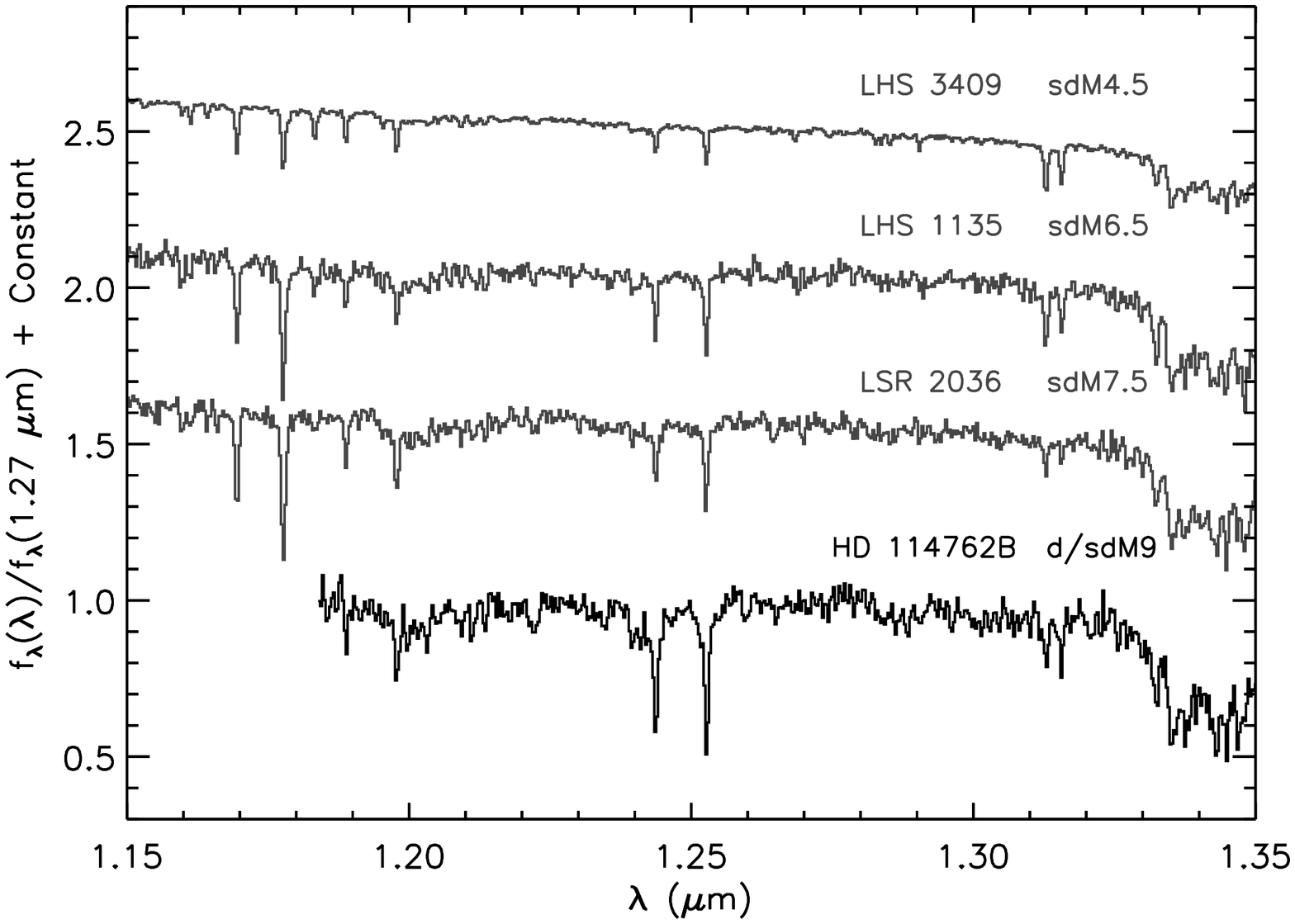}}
  \caption{$J$-band spectra of increasingly later-type metal-poor objects compared to our OSIRIS spectrum of HD 114762B.  The data for LHS 3409, LHS 1135, and LSR 2036+5059 are from \citet{Cushing:2006p562} taken with SpeX  in SXD mode at IRTF.  Spectra are normalized to 1.27 $\mu$m and are shifted by a constant.  HD 114762B is consistent with a spectral type later than LSR 2036+5059 (sdM7.5) based on the depth of the 1.2436 and 1.2526 $\mu$m \ion{K}{1} lines and the strength of the 1.4 $\mu$m H$_2$O steam band.  The spectrum of HD 114762B has been Gaussian smoothed to the same resolution as the other data.  \label{f6} \\}
\end{figure}

The best overall fit appears to be to the d/sdM9 object 2MASS 00412179+3547133 (\citealt{Burgasser:2004p574}).  We therefore adopt a spectral type of d/sdM9 $\pm$ 1 for HD 114762B.  In Figure \ref{f5} we compare 2MASS 0041+35 to the shifted SpeX spectrum.  The agreement is satisfactory over the $J$ and $K$ bands, but the $H$ band is significantly lower.  This reduced $H$ band flux  would only worsen the agreement between the shifted spectrum and the spectra from Figure \ref{f4}; we therefore suspect that this lowered flux is an artifact of shifting the segments of the SpeX data to published photometry which had large uncertainties (color errors were $\sim$ 0.15 mags).  In addition, we note that the d/sdM9 spectral type for HD 114762B is consistent with the sequence of medium-resolution subdwarf spectra shown in Figure \ref{f6} based on the strength of the 1.4 $\mu$m steam band and the strengths of the 1.2436 and 1.2526 $\mu$m \ion{K}{1} lines.

Another line of evidence independently bolsters the metal-poor nature of HD 114762B.  This object has a blue $J$ -- $K_\mathrm{S}$ color (0.73 mag) compared to normal dwarfs with (optical) spectral types later than $\sim$ M6, which have $J$ -- $K_\mathrm{S}$ values $\gtrsim$ 1.0 mags (see, e.g., \citealt{Kirkpatrick:2008p10975}).  Blue near-infrared colors may be produced  through strong CIA H$_2$ caused by a reduced metallicity.  It can also originate from  thin or large grain condensate clouds (\citealt{Burgasser:2008p3725}).  Clouds begin to appear in the latest M dwarfs (\citealt{Jones:1997p17642}; \citealt{Allard:2001p14776}), but their influence on the near-infrared spectral slope only becomes significant in L dwarfs, as is evident by the increased spread in $J$ -- $K_\mathrm{S}$ color in that spectral class.  The blue near-infrared colors observed in HD 114762B are therefore indicative of a reduced metallicity.

\begin{figure*}
  \plotone{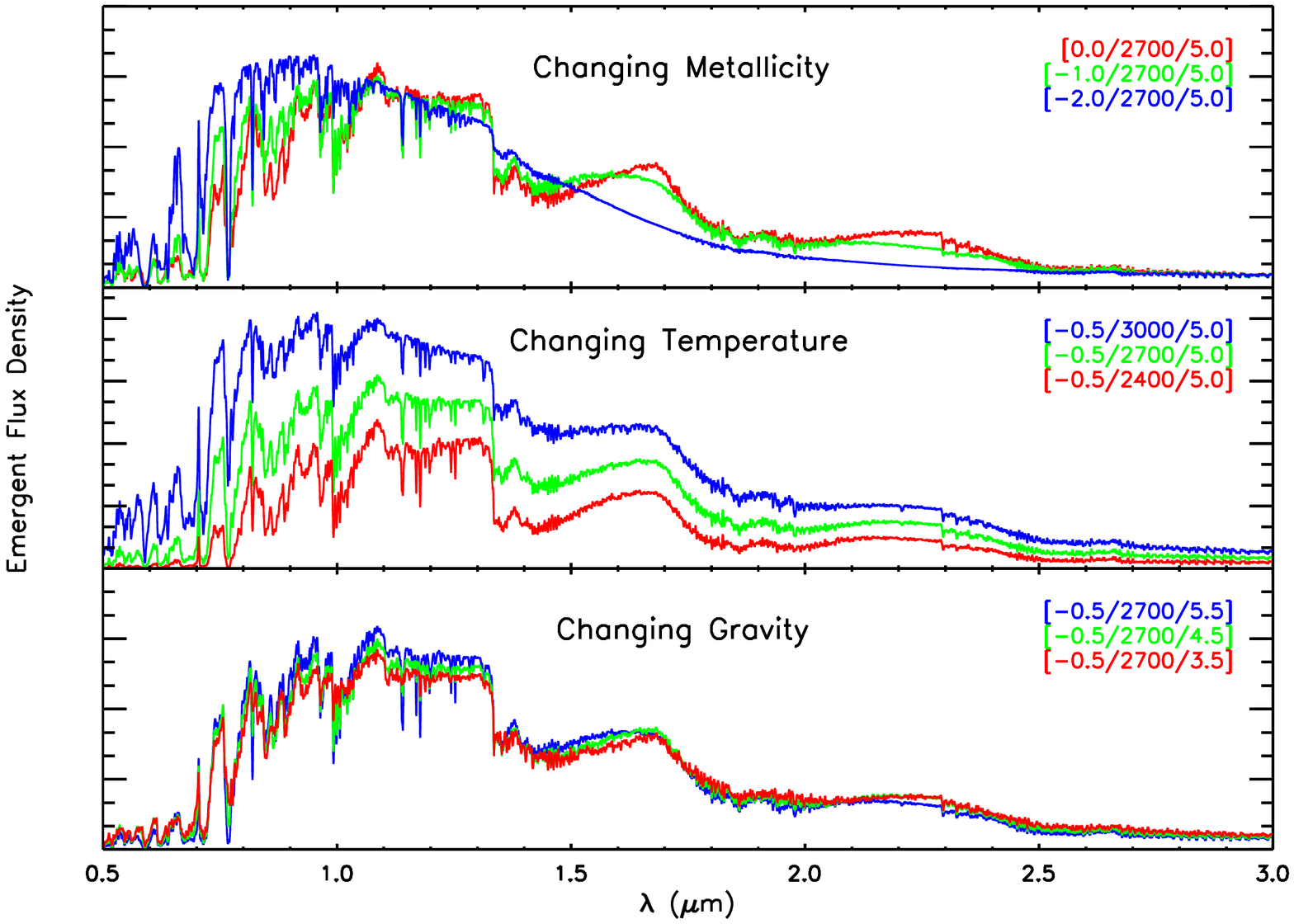}
  \caption{The PHOENIX/$GAIA$ v.2.6.1 atmospheric models (\citealt{Brott:2005p301}) showing the variations in emergent flux density with metallicity (top), temperature (middle), and surface gravity (bottom).  The result of a reduced metallicity includes a suppression of $H$ and $K$-band flux from collision-induced H$_2$ absorption, producing bluer near-infrared colors.  The model notation is as follows:  [[M/H], $T_\mathrm{eff}$ (K), log $g$].  \label{f7} \\ } 
\end{figure*}

\subsection{Physical Parameters From Atmospheric Models}

There is no standard method for fitting a coarsely-sampled grid of synthetic spectra (e.g., $\Delta$[M/H] $\sim$ 0.5, $\Delta$$T_\mathrm{eff}$ $\sim$ 100 K, and $\Delta$log $g$ $\sim$ 0.5) to observations.  General techniques for fitting models to data usually rely on minimizing (some form of) the $\chi^2$ statistic, equal to $\sum_i$ ($f_i$ -- $\mathcal{M}_i$)$^2$/$\sigma_i^2$, where $f_i$, $\mathcal{M}_i$, and $\sigma_i$ are the data, the model prediction, and the measurement error at  point $i$, respectively.  When the random variable ($f_i$ -- $\mathcal{M}_i$)/$\sigma_i$ is a standard normal deviate, the $\chi^2$ statistic approximates a $\chi^2$ distribution.  In that case confidence regions for the best-fitting parameters may be approximated by calculating the $\chi^2_\mathrm{min}$-plus-constant value (e.g., \citealt{Press:2007p13558}).  

Atmospheric models have known systematic errors in the form of incomplete opacity sources, in which case the $\chi^2$ method of error analysis may not be applicable.  Thus the question arises of how correct the models must be for this $\chi^2$ technique to remain reliable.  Small deviations from ``truth'' in the model will have little influence on the resulting confidence regions, but large deviations will make this method  more and more incorrect.  One way to assess this applicability is to study the normality of the distribution of deviates ($f_i$ -- $\mathcal{M}_i$)/$\sigma_i$ for the best-fitting model.  A near-perfect model will have deviates that are approximately normally distributed, but an inadequate model will have non-normally distributed deviates.  Note that the condition of normally distributed deviates does not guarantee that the model is good, but a good model should have normally distributed deviates.  A test for normality of the deviates may therefore serve as a useful assessment of the correctness of a best-fitting model.  Reduced $\chi^2$ values are also used as a goodness-of-fit measure, but they rely on an accurate determination of measurement errors, a task that is difficult to correctly perform.

\subsubsection{Method}

In this study we first determine the synthetic spectrum that best fits the data using a Monte Carlo approach described below.  We then make use of two parameters to determine the goodness-of-fit: the reduced $\chi^2$ value and the significance level of the D'Agostino-Pearson normality test (\citealt{DAgostino:1990p11765}) applied to the best-fitting model deviates.  The latter test compares the skewness and kurtosis of a distribution to that of a normal distribution.  If the deviates are normally distributed and the reduced $\chi^2$ value is low (we impose a cutoff of 2.0) then the model is deemed ``good'' and approximately Gaussian confidence limits can be derived using the aforementioned $\chi^2_\mathrm{min}$-plus-constant method.  If these criteria are not satisfied then we simply report the physical parameters of the top few best-fitting models without errors.

We use the $GAIA$ grid of synthetic spectra, derived from the PHOENIX stellar atmosphere code ($GAIA$ version 2.6.1, \citealt{Brott:2005p301}; \citealt{Hauschildt:1999p81}), as models for our observations.  We made use of 4030 models  with zero $\alpha$-element enrichment compared to solar values covering the cool stellar and substellar regime: --4.0 $\le$ [M/H] $\le$ 0.5 ($\Delta$[M/H] = 0.5), 2000 K $\le$ $T_\mathrm{eff}$ $\le$ 5000 K ($\Delta$$T_\mathrm{eff}$ = 100 K), --0.5 $\le$ log $g$ $\le$ 5.5 ($\Delta$log $g$ = 0.5).  Examples of these models over a range of metallicities, temperatures, and surface gravities are presented in Figure \ref{f7}.

Our approach for fitting the synthetic spectra to the observations makes use of a Monte Carlo method previously used by several authors (e.g., \citealt{Saumon:2006p10455}, \citealt{Cushing:2008p2613}).  The $GAIA$ models were Gaussian smoothed to the resolution of the SpeX and OSIRIS data and were resampled onto the same wavelength grids.  The synthetic spectra provide the emergent flux density at the surface of a star and have to be scaled to the flux density observed at Earth.  For each model $k$ we compute the scaling factor $C_k$ by minimizing the $\chi^2$ statistic

\begin{equation}
\chi^2 = \sum_{i=1}^{n}{\left( \frac{f_i - C_k \mathcal{E}_{k,i}} { \sigma_i} \right)^2},
\end{equation}

\noindent where $f_i$ and $\sigma_i$ are the observed flux density and the measurement error, respectively, $\mathcal{E}_{k,i}$ is the emergent flux density for model $k$ and pixel $i$, and $n$ is the number of data points used in the fit.  The scaling factor is given by

\begin{equation}
C_k = \frac{ \sum f_i \mathcal{E}_{k,i}/\sigma_i^2}{\sum \mathcal{E}_{k,i}^2/\sigma_i^2}
\end{equation}

\noindent and is related to the stellar radius $R$ and the distance $d$ to the source by $C_k$ = $(R/d)^2$.  

The $GAIA$ model with the lowest $\chi^2$ value may change for different observations of the same source as a result of random noise in the data.   To account for this uncertainty we generated a large set of simulated data and then determined the best-fitting model for each new spectrum.  The simulated data were created by adding noise to the original spectrum using the measurement errors.  To account for the uncertainty in the flux calibration, which is important for the error in the model scaling factor $\sigma_{\overline{C}_k}$, the artificial spectrum was adjusted by a flux calibration scaling factor drawn from a Gaussian distribution with a mean $\overline{C}_\mathrm{fc}$ and a standard deviation $\sigma_{\overline{C}_\mathrm{fc}}$.  Following the approach of \citet{Cushing:2008p2613}, the 20 models  with the lowest original $\chi^2$ values were then refit to the artificial data sets for 10,000 trials of simulated spectra.  We adopted the model with the highest fraction of $\chi^2_\mathrm{min}$ values as the best-fitting model; we refer to that fraction as $f_\mathrm{MC}$, the ``Monte Carlo fraction'' (\citealt{Cushing:2008p2613}).  A mean scaling factor $\overline{C}_k$ was calculated by averaging the individual $C_k$ values for trials where the model with the highest resulting Monte Carlo fraction had the lowest $\chi^2$ value.  The scaling factor uncertainty $\sigma_{\overline{C}_k}$ represents the standard deviation of the distribution of $C_k$ values.

\begin{figure}
  \resizebox{3.45in}{!}{\includegraphics{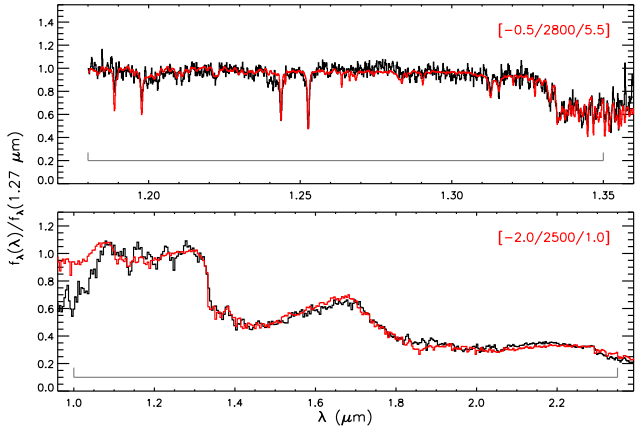}}
  \caption{Results of fitting the OSIRIS (top) and SpeX (bottom) spectra.  The best-fitting $GAIA$  model to the 1.18-1.35 $\mu$m region of the OSIRIS spectrum was [--0.5/2800/5.5].  The best-fitting model to the SpeX spectrum from 1.00-2.35 $\mu$m was [--2.0/2500/1.0].  The metallicity from fitting the OSIRIS spectrum ([M/H] = --0.5) is consistent with that of the primary star HD 114762A ([Fe/H] = --0.70).  The gray regions at the bottom indicate the sections of the observations that were used in the fit.  \label{f8} } 
\end{figure}

\begin{figure}
  \resizebox{3.45in}{!}{\includegraphics{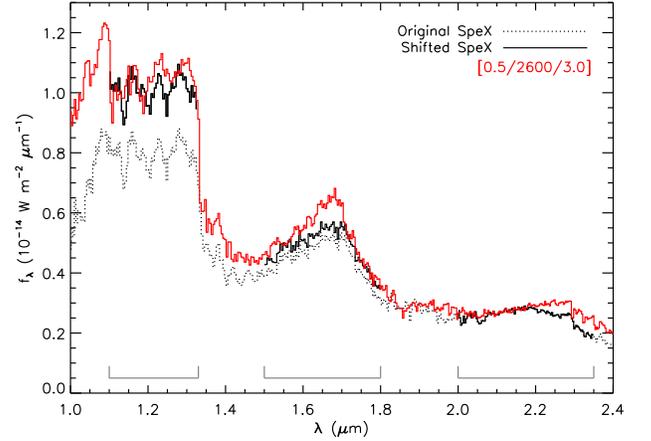}}
  \caption{The best-fitting model (red) to the shifted SpeX spectrum (solid black) compared to the original unshifted SpeX spectrum (dotted black).   The resulting metallicity ([M/H] = +0.5) is inconsistent with the value of the primary star ([Fe/H] = --0.70). The gray regions at the bottom indicate the sections of the observations that were used in the fit. \label{f9} } 
\end{figure}

\subsubsection{Results}

\begin{figure}
  \resizebox{3.45in}{!}{\includegraphics{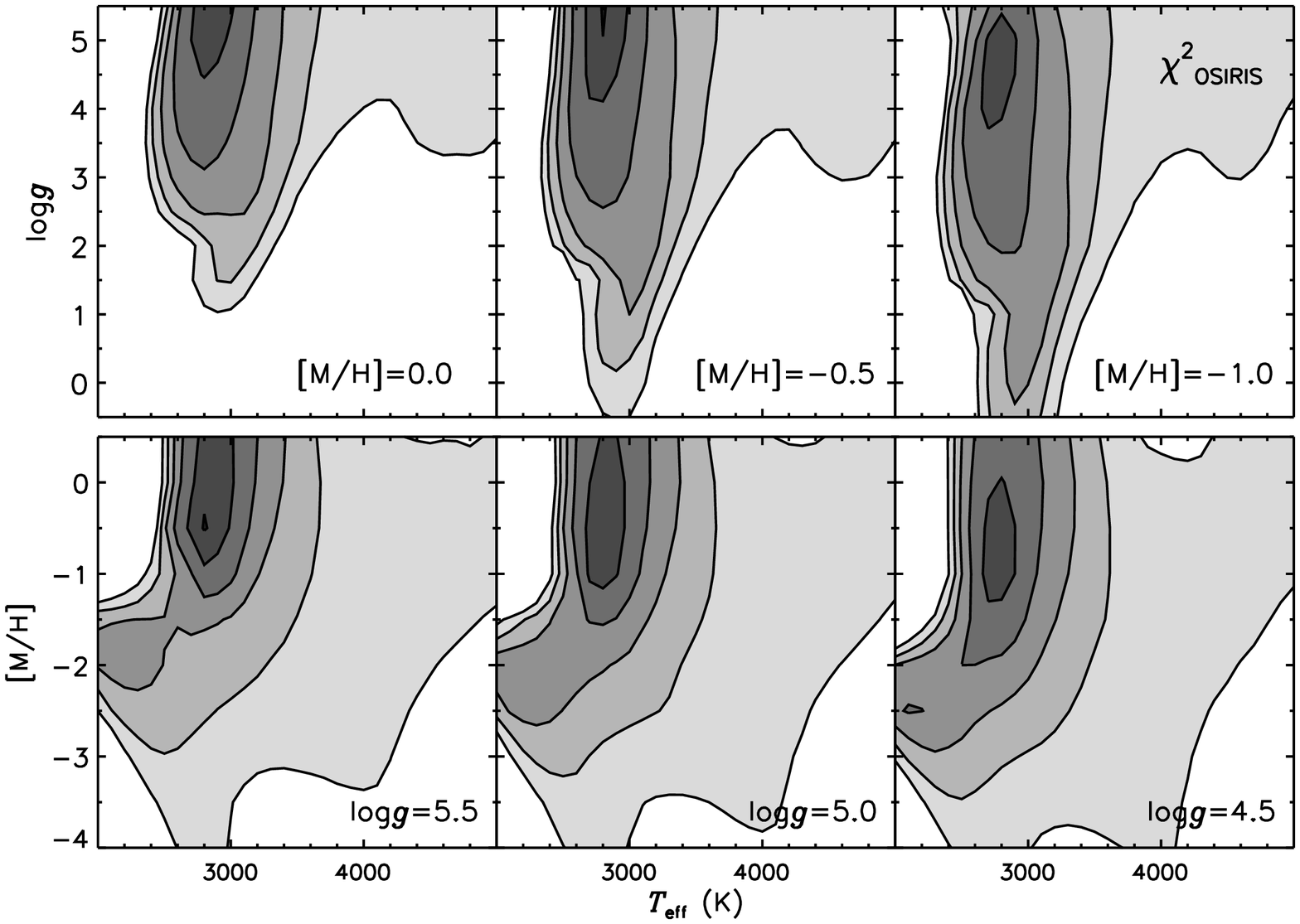}}
  \caption{Contour plots of $\chi^2$ values from fitting the $GAIA$ synthetic spectra to our OSIRIS spectrum.  The top row shows slices in metallicity near the global minimum; the bottom row shows slices in surface gravity.  The global minimum is located at [M/H] = --0.5, $T_\mathrm{eff}$ = 2800 K, log $g$ = 5.5.   The contour levels represent 1.02, 1.3, 1.9, 2.5, 3.1, and 3.7 times the $\chi^2_\mathrm{min}$ value, where $\chi^2_\mathrm{min}$ = 8222.7 (dof = 1128).  Although confidence regions of contours cannot be computed using the $\chi^2_\mathrm{min}$-plus-constant method (see $\S$ 3.2.1), the overall shapes are useful aids for visualizing how constrained each parameter is.  Local minima can  also be easily spotted in this way.  In this case the temperature is well constrained, but the surface gravity and metallicity are not.  \label{f10}} 
\end{figure}

\begin{figure}
  \resizebox{3.45in}{!}{\includegraphics{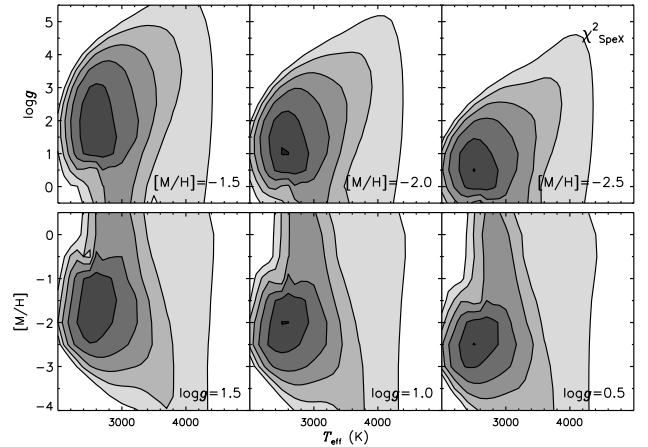}}
  \caption{Contour plots of $\chi^2$ values from fits to our SpeX spectrum (similar to Figure \ref{f10}).  The global minimum is located at [M/H] = --2.0, $T_\mathrm{eff}$ = 2500 K, log $g$ = 1.0.  The contours represent values of 1.02, 2.0, 4.0, 6.0, 8.0, and 10.0 times the $\chi^2_\mathrm{min}$ value, where $\chi^2_\mathrm{min}$ = 2324.4 (dof = 359). \label{f11}} 
\end{figure}

Accurate atmospheric models should yield similar best-fitting physical properties for fits to observations of the same target at different wavelengths and spectral resolutions.  We fit the $GAIA$ synthetic spectra to the OSIRIS spectrum (from 1.18-1.35 $\mu$m) and to the SpeX spectrum (from 1.00-2.35 $\mu$m; Figure \ref{f8}).  We also fit the shifted SpeX spectrum (Figure \ref{f9}, see $\S$ 2.4) as well as the individual near-infrared bandpasses of the original SpeX spectrum (from 1.10-1.33 $\mu$m for $J$, 1.50-1.80 $\mu$m for $H$, and from 2.05-2.35 $\mu$m for $K$) to study the wavelength dependence of the fits.  Our Monte Carlo fitting of the shifted SpeX spectrum incorporated measurement errors from both the observed spectrum and the published photometry.  For the shifted SpeX spectrum, each spectral segment was adjusted by a scaling factor drawn from a normal distribution with a standard deviation equal to the error in the published photometry.  Then measurement errors were drawn and added back to the data.  Results from the fits are presented in Table \ref{tabfit}.  In Figures \ref{f10} and \ref{f11} we present contour plots of slices of the $\chi^2$ cubes near the best-fitting models for fits to the entire OSIRIS and SpeX spectra.

\begin{deluxetable*}{lcccccc}
\tabletypesize{\scriptsize}
\tablewidth{0pt}
\tablecolumns{7}
\tablecaption{Best-Fit PHOENIX/$GAIA$ Models \label{tabfit}}
\tablehead{
\colhead{Spectral Region}   &    \colhead{}     &    \colhead{}    &  \colhead{}         & \colhead{}  &  \colhead{}   & \colhead{}  \\
\colhead{Used in Fit}   &    \colhead{[M/H]}     &    \colhead{$T_\mathrm{eff}$ (K)}    &  \colhead{log $g$}         & \colhead{$f_\mathrm{MC}$}  &  \colhead{$p$}   & \colhead{$\chi^2_\mathrm{min}/\nu$} 
}
\startdata
\cutinhead{OSIRIS}
1.18 - 1.35 $\mu$m          & \bf{--0.5}  & \bf{2800} &\bf{5.5}  &  \bf{0.998}\tablenotemark{a} & $<$ 10$^{-3}$ & 7.29  \\
Absorption Lines\tablenotemark{b} &  \bf{--0.5}  & \bf{2600} & \bf{5.0}   &   \bf{0.824}\tablenotemark{a} & $<$ 10$^{-3}$ & 5.58 \\
                                                                                              &  0.0          & 2500          & 5.0     &  0.176  &  $<$ 10$^{-3}$  &  5.69 \\  
\cutinhead{SpeX} 
1.00 - 2.35 $\mu$m  & \bf{--2.0}  & \bf{2500} & \bf{1.0}  &   \bf{0.500}\tablenotemark{a}& $<$ 10$^{-3}$ &6.47  \\ 
                                                                                                                & --2.5         &  2500       &   0.5       &      0.343  & $<$ 10$^{-3}$ &  6.49  \\
                                                                                                                & --2.0         &  2600       &   1.0       &      0.157   & $<$ 10$^{-3}$  &  6.53 \\
1.00 - 1.35 $\mu$m  &  \bf{--1.5}  & \bf{2600} & \bf{3.5}   &  \bf{ 0.697}\tablenotemark{a} & 0.104 & 3.82  \\
                                                    & --1.0   & 2600 & 4.0&   0.231 & 0.305 & 3.89 \\
1.50 - 1.80 $\mu$m  & \bf{--1.5}  & \bf{2600} & \bf{1.5}    & \bf{0.422}\tablenotemark{a}  & 0.911 & 5.87 \\
                                                                                     & --1.0  & 2600 & 2.0   &   0.378 & 0.921 & 5.87 \\ 
                                                                                     & --3.0  & 2700 & --0.5   &   0.174 & 0.902 & 5.99 \\ 
2.05 - 2.35 $\mu$m  & \bf{--0.5}  & \bf{2500} & \bf{5.0}      &   \bf{0.527}\tablenotemark{a}& $<$ 10$^{-3}$ & 1.47 \\
                                                    & --0.5   &  2600 & 5.0   &   0.287& $<$ 10$^{-3}$ & 1.48 \\
                                                     
\cutinhead{SpeX Shifted to Match Published Photometry\tablenotemark{c}}
$J$/$H$/$K$ bands  &  \bf{0.5}  &  \bf{2600}  &  \bf{3.0}       &    \bf{0.392}& $<$ 10$^{-3}$ & 8.09 \\  
  &  0.5  &  2700  &  3.5   &  0.130 \tablenotemark{a} & $<$ 10$^{-3}$ & 7.20 \\

\enddata
\tablecomments{Synthetic spectra are from the PHOENIX/$GAIA$ v.2.6.1 model atmosphere code (\citealt{Brott:2005p301}).  Column 1 lists the region of the spectrum used in the fit.  The resulting metallicities, effective temperatures, and surface gravities are listed in Columns 2, 3, and 4, respectively.  The Monte Carlo fraction ($f_\mathrm{MC}$) is listed in Column 5 ($\S$3.2.1).  Models with $f_\mathrm{MC}$ $\ge$ 0.1 are reported, and the model with the highest fraction is shown in bold.  A D'Agostino-Pearson normality test is applied to the deviates for each model.   The significance level $p$ of the null hypothesis, that the deviates are normally distributed, is listed in Column 6.  The null hypothesis is rejected for values of $p$ below 0.005.  Reduced $\chi^2$ values for fits to the \emph{observed data} rather than the simulated data are listed in Column 6, where $\nu$ = $n$ -- $m$ --1 for $n$ data points and $m$ parameters used in the model.  Note that $m$ = 4 for all fits except for fits to the absorption lines, in which case $m$ = 7 ($T_\mathrm{eff}$, [M/H], log $g$, and one scaling factor for each of the four wavelength ranges).}

\tablenotetext{a}{This model had the lowest $\chi^2$ value in a fit to the observed data.}

\tablenotetext{b}{Five lines were used: Fe I (1.1886/1.1887 $\mu$m; blended), K I (1.2436 $\mu$m), K I (1.2526 $\mu$m), Al I (1.3127 $\mu$m), Al I (1.3154 $\mu$m).}

\tablenotetext{c}{Shifted to match photometry from \citet{Patience:2002p186}. The $\chi^2$ fit was performed to the $J$, $H$, and $K_\mathrm{S}$ bands, skipping over the steam bands.  See $\S$2.4 for details. \\ }

\end{deluxetable*}

The resulting metallicities and surface gravities vary greatly among the best-fitting models from Table \ref{tabfit}, but there is good agreement in the derived effective temperatures (2500-2800 K).  The fitting results that suggest a low surface gravity for this object are inconsistent with the known high surface gravities of old low-mass stars and brown dwarfs (\citealt{LopezMorales:2007p4016}; see Figure 2 of \citealt{Jao:2008p6434} for a graphical version of the data).  The fit to the OSIRIS spectrum resulted in a best-fitting model with [M/H] = --0.5 and log $g$ = 5.5, in accord with the metallicity of the primary\footnote{We do not differentiate between [Fe/H] and [M/H] for our tests, although we note that \citet{Valenti:2005p11833} derive a value of [M/H] = --0.52 for HD 114762A.  This difference does not affect the analysis nor the results, however.} and the high surface gravities of late-type objects.  Reduced $\chi^2$ values are generally quite high and the significance levels ($p$) from the D'Agostino-Pearson normality tests are mostly low, indicating that the best-fitting models do not match the observations well.  A visual inspection confirms this, although visually the OSIRIS fit appears to be quite good (``chi-by-eye'').  If the $p$ value is lower than 0.005 then we reject the null hypothesis that the deviates are normally distributed.  In no case were both $p$ $>$ 0.005 and $\chi^2_\mathrm{min}/\nu$ $<$ 2.0 so a formal error analysis on the fitted physical parameters is not performed.

\begin{deluxetable*}{ccccccc}
\tabletypesize{\scriptsize}
\tablewidth{0pt}
\tablecolumns{7}
\tablecaption{PHOENIX/$GAIA$ Model Fits to Published Late-Type Mild Subdwarf SpeX Spectra \label{templates}}
\tablehead{
\colhead{Object\tablenotemark{a}}    &   \colhead{[M/H]}         &   \colhead{$T_\mathrm{eff}$ (K)}          & \colhead{log $g$}       & \colhead{$f_\mathrm{MC}$} &  \colhead{$p$}   & \colhead{$\chi^2_\mathrm{min}/\nu$}   
 }
\startdata
\cutinhead{Fit from 0.80-2.35 $\mu$m}
2MASS 0115+31 (d/sdM8)     & \ \ 0.5  & 2600 & 4.0  &  0.999        & $<$ 10$^{-3}$ & 6.95    \\
2MASS 1556+13 (d/sdM8)   & \ \ 0.0  & 2500 & 3.0  & 1.000     & $<$ 10$^{-3}$  &  9.24   \\
2MASS 1559--03 (d/sdM8)   &  \ \ 0.5  & 2600 & 3.5  &  0.814      & $<$ 10$^{-3}$  & 7.24      \\
2MASS 0041+35 (d/sdM9)  &  --3.0  & 2300 & 0.0  &  0.972       & $<$ 10$^{-3}$ &  15.04   \\

\cutinhead{Fit from 1.00-2.35 $\mu$m}
2MASS 0115+31       &  \ \ 0.0  &2700 & 4.5  &  1.000  & $<$ 10$^{-3}$     & 3.84   \\
2MASS 1556+13   &  --1.5  & 2600 & 2.0  &  0.959   & $<$ 10$^{-3}$    & 4.66  \\
2MASS 1559--03  &  \ \ 0.5  & 2600 & 4.5  &  0.889    & $<$ 10$^{-3}$    & 6.68   \\
2MASS 0041+35   &  --1.5  & 2600 & 3.0  &  0.920     & $<$ 10$^{-3}$  &  5.86  \\

\cutinhead{Fit from 1.00-1.35 $\mu$m}
2MASS 0115+31   &  --1.5  & 2600 & 3.0  &  0.231        & 0.206  & 3.47    \\
2MASS 1556+13   &  --2.5  & 2600 & 1.5  & 0.398     & 0.484  &  3.21   \\
2MASS 1559--03 &  --1.5  & 2600 & 2.5  &  0.637      & 0.866  & 3.28      \\
2MASS 0041+35 &  --1.0  & 2300 & 4.0  &  0.373       & 0.404 &  5.26   \\

\cutinhead{Fit from 1.50-1.80 $\mu$m}
2MASS 0115+31    &  --0.5  & 2700 & 4.0  &  0.822        & 0.060  & 3.25    \\
2MASS 1556+13    &  --2.0  & 2600 & 1.0  & 0.852     & 0.920  &  4.68   \\
2MASS 1559--03   &  \ \ 0.0  & 2700 & 4.5  &  0.429      & 0.664  & 6.48      \\
2MASS 0041+35 &  \ \ 0.0  & 2500 & 5.0  &  0.782       & 0.702 &  5.38   \\

\cutinhead{Fit from 2.05-2.35 $\mu$m}
2MASS 0115+31   &  \ \ 0.0  & 2800 & 5.0  &  0.498        & 0.865  & 1.67    \\
2MASS 1556+13   &  \ \ 0.5  & 2800 & 5.5  & 0.855     & 0.022  &  2.17   \\
2MASS 1559--03   &  --0.5  & 2700 & 4.0  &  0.370      & 0.708  & 4.24      \\
2MASS 0041+35  &  --1.0  & 2700 & 4.5  &  0.487       & $<$ 10$^{-3}$ &  2.56   \\

\enddata
\tablenotetext{a}{The full names of the objects are 2MASS 01151621+3130061, 2MASS 15561873+1300527,  2MASS 15590462--0356280, and 2MASS 00412179+3547133.}
\end{deluxetable*}

\begin{figure}
  \resizebox{3.45in}{!}{\includegraphics{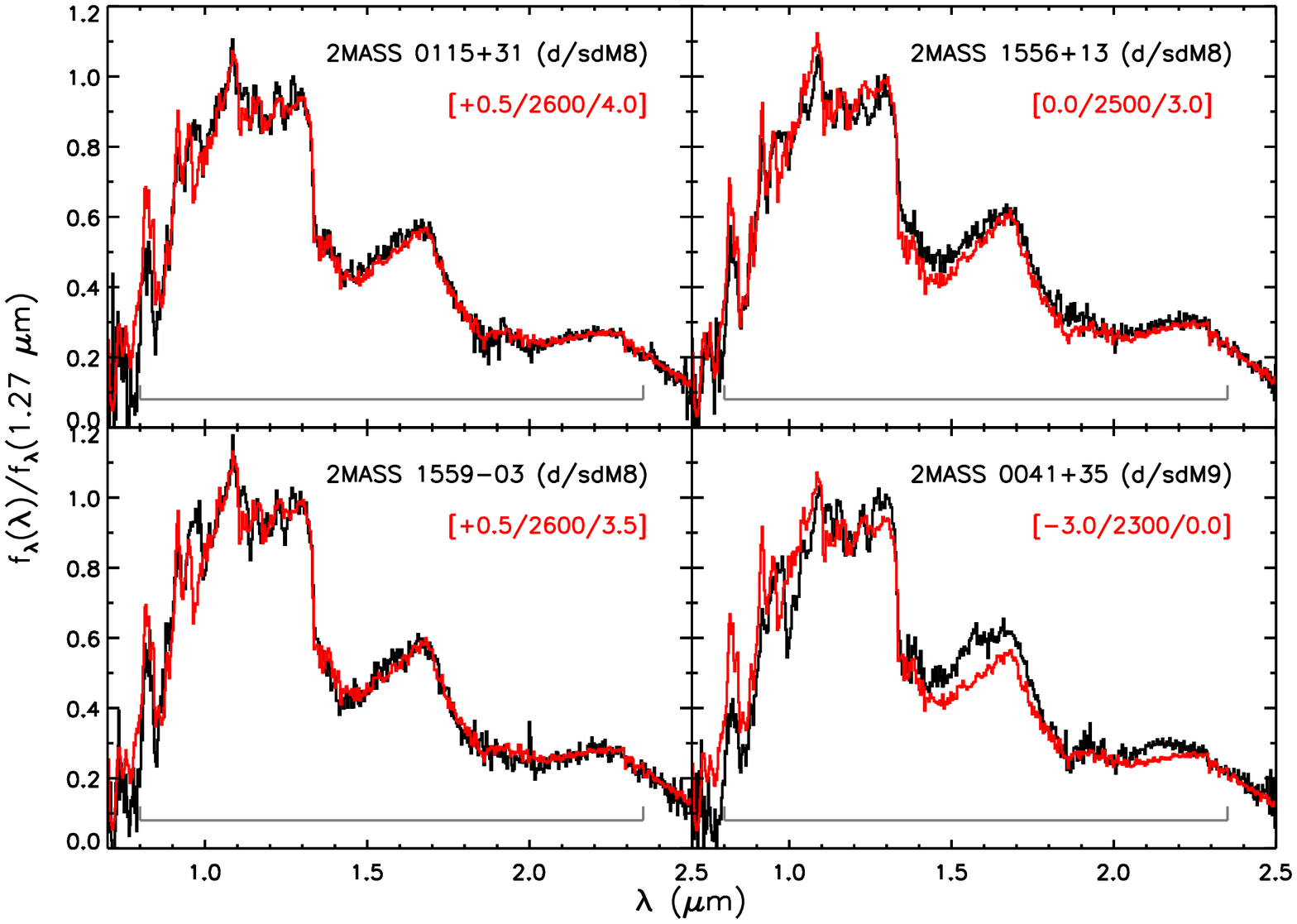}}
  \caption{Spectral fits to SpeX/prism data for objects with similar spectral types to HD 114762B (from \citealt{Burgasser:2004p574}).  The metallicities and surface gravities from the best-fitting models vary considerably among the objects for fits to the 0.80-2.35 $\mu$m region. The full names of the mild subdwarfs are 2MASS J01151621+3130061, 2MASS J15561873+130027, 2MASS J15590462--0356280, and 2MASS J00412179+3547133.  The gray lines at the bottom indicate the sections of the observations that were used in the fit. \label{f12} } 
\end{figure}

Fits to the SpeX spectrum yielded a large range of metallicities (--2.0 $\le$ [M/H] $\le$ --0.5 dex) and surface gravities (1.0 $\le$ log $g$ $\le$ 5.0).  It is unclear whether these discrepancies result from a failure of the $GAIA$ models or originate from the SpeX data themselves, e.g. as a result of the atypical spectral extraction technique (\S 2.2.2).  To address this question we fit the models to published near-infrared spectra of four ultracool subdwarfs with similar spectral types to that of HD 114762B.  If the models predict mutually consistent values with only small variations in each parameter then that would indicate that it is our SpeX spectrum that may be skewing the results of the fits.  We examined the spectra of three mildly metal-poor d/sdM8 objects (2MASS 01151621+3130061, 2MASS 15561873+1300527, 2MASS 15590462-0356280; NIR spectral types) and one d/sdM9 object (2MASS 00412179+3547133; NIR spectral type) from \citet{Burgasser:2004p574}, made available through the SpeX Prism Spectral Library.   We used the measurement errors from our SpeX observations scaled to the square root of the ratio of the respective integration times to determine $f_\mathrm{MC}$ values\footnote{Initially we used the errors provided in the SpeX Prism Spectral Library which were derived from the observations, but reduced chi-squared values were quite high, indicating that the errors were likely underestimated.}.  We first fit the entire 0.80-2.35 $\mu$m range of the prism spectra, presented in Figure \ref{f12}.  The resulting metallicities were 0.5, 0.0, 0.5, and --3.0 dex for the four (presumably) mildly metal-poor objects. To test the wavelength dependence of the predictions we also fit the mild subdwarfs from 1.00-2.35 $\mu$m and over the individual $J$, $H$, and $K$ bandpasses (Table \ref{templates}).  The resulting metallicities and surface gravities varied considerably among the four objects.  Fits to different spectral regions of the same object also produced highly discrepant results.  The effective temperatures were relatively constant, however, with most fits giving 2500 K or 2600 K.  The overall inconsistency implies that \emph{the metallicities and surface gravities resulting from $\chi^2$ fits to low-resolution near-infrared spectra of ultracool subdwarfs are probably not reliable}.

\begin{figure*}
  \plotone{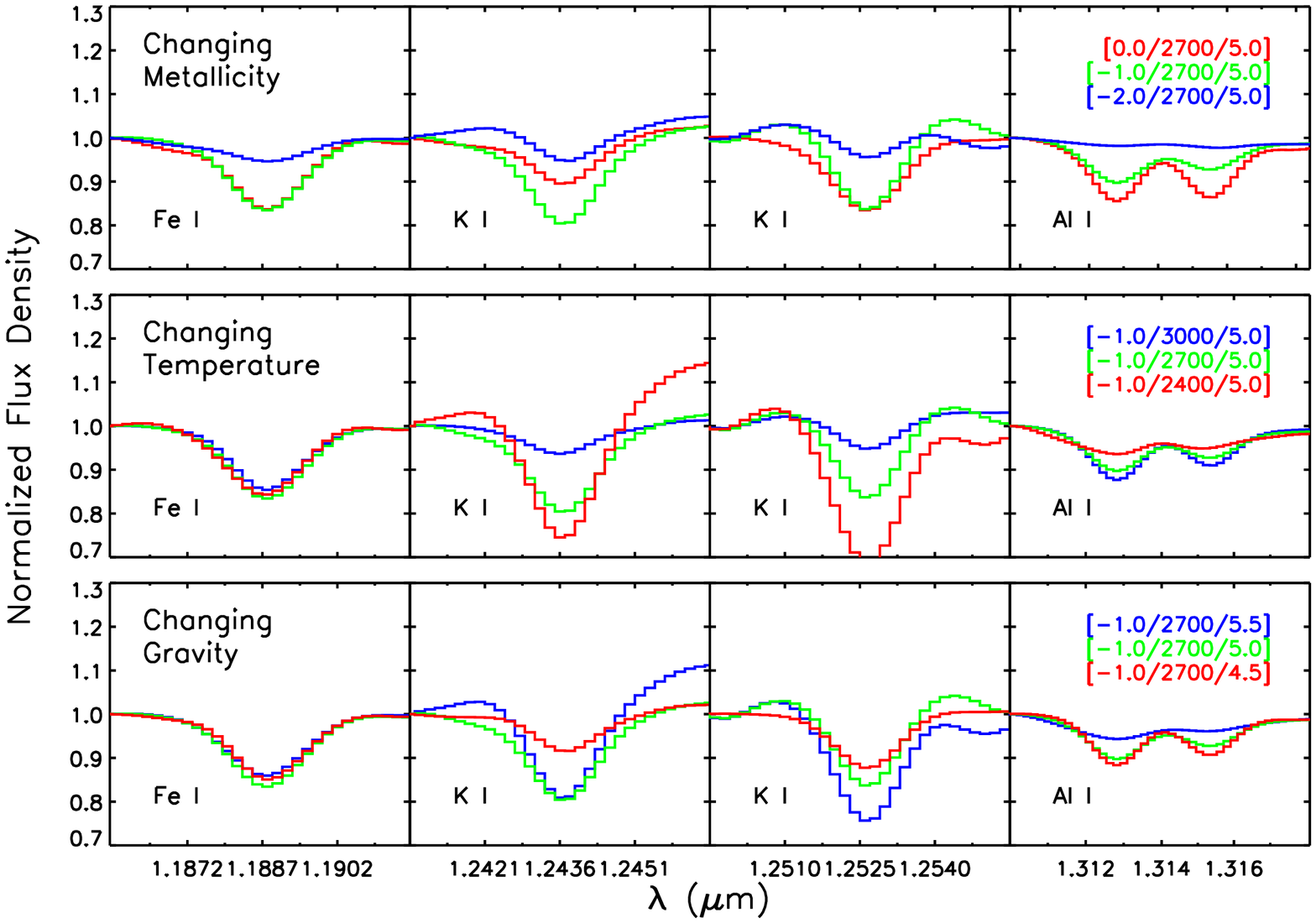}
  \caption{Relative strengths of the $J$-band absorption lines in the $GAIA$ model atmospheres for changing physical parameters.  The depth of the lines are sensitive to changes in metallicity (top), temperature, (middle), and surface gravity (bottom).  The combination of these lines provides a good estimator of all three parameters when fitting synthetic spectra to observations.  The models are smoothed for better rendering.   \label{f13} } 
\end{figure*}

\begin{figure}
  \resizebox{3.5in}{!}{\includegraphics{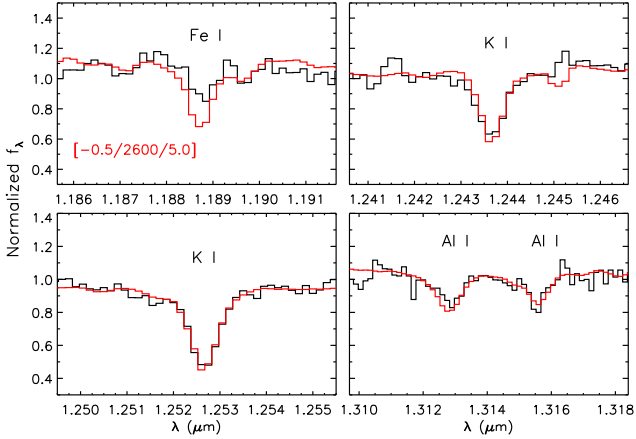}}
  \caption{Results from fitting the OSIRIS data to the absorption lines.  The metallicity and surface gravity are consistent with the metallicity of the primary star ([Fe/H] = --0.70) and the high surface gravity of late-type objects.  The effective temperature is consistent with that predicted by the CB97 evolutionary models.  Note that the depth to the 1.1887 $\mu$m \ion{Fe}{1} line is overestimated in the model. \label{f14} } 
\end{figure}

We also fit the metal-sensitive absorption lines in our OSIRIS spectrum to determine whether this fitting technique was superior to fitting the entire $J$ band spectral region, which emphasizes the overall shape of the spectrum rather than the strengths of the individual absorption lines.  We identified prominent lines in the OSIRIS observations which were sensitive to metallicity, temperature, and surface gravity in the $GAIA$ models (see Figure \ref{f13}).  Five lines were chosen: \ion{Fe}{1} (1.1887 $\mu$m), \ion{K}{1} (1.2436 $\mu$m), \ion{K}{1} (1.2526 $\mu$m), \ion{Al}{1} (1.3127 $\mu$m), and \ion{Al}{1} (1.3154 $\mu$m).  We fit these lines in the OSIRIS data using a fitting range of $\pm$ 30 \AA \  from the center of each absorption line, except for the aluminum doublet, for which the range was $-$30 \AA \  from the 1.3127 $\mu$m line to $+$30 \AA \  from the 1.3154 $\mu$m line.  The data were normalized to unity at the shortest wavelength of the fitting range for each line and a separate scaling factor was computed for each of the four spectral segments.  Each model was then fit independently to the four regions and the resulting $\chi^2$ values were summed to produce a final $\chi^2$ value for that model.  The parameters from the best-fitting models using this technique are similar to those for fits to the entire OSIRIS $J$ band spectrum (Figure \ref{f14}), although in this case the effective temperature is reduced by 200 K to 2600 K\footnote{We also tried fitting the models using the pseudo-equivalent widths of the model absorption lines relative to those of the data.  We calculated a chi squared statistic
\begin{equation}
\chi^2_{EW} = \sum_{ l = 1}^{5}  \frac{ (EW_{f, l}  -  EW_{\mathcal{M}, l})^2 } {  EW_{\mathcal{M},  l}  }   
\end{equation}
over lines $l = $ 1 to 5.  Here $EW_{f, l}$ is the pseudo-equivalent width of the observed data for line $l$, and $EW_{\mathcal{M}, l}$ is the pseudo-equivalent width of the model line.  A Monte Carlo simulation was used to determine Monte Carlo fractions. $f_\mathrm{MC}$ values were quite low, indicating that this method did a poor job of distinguishing among the best-fitting models.  The results were also dependent on the way the pseudo-equivalent widths were calculated, as the pseudo-continuum is far from smooth.  This method seems to be inferior to $\chi^2$ fits over the absorption lines.}.

The effective temperatures of the best-fitting $GAIA$ models (2500-2800 K) are systematically hotter than the temperatures of other late M-type objects which have been derived using other techniques.  For example, \citet{Gautier:2007p19058} use  24 $\mu$m photometry from \emph{Spitzer Space Telescope} to derive effective temperatures of M dwarfs using the infrared flux method.  They find effective temperatures of 2150-2450 K for spectral types between M8 and M9.5.  Similarly, \citet{Golimowski:2004p15703} derive effective temperatures between 2000-2525 for spectral types between M8.5 and M9.5 based on $L_\mathrm{Bol}$-$T_\mathrm{eff}$ relations from evolutionary models.   The hotter effective temperatures we inferred from this study may have two (non-mutually exclusive) causes: systematic errors in the models or different temperature scales caused by the metal-poor nature of HD 114762B.  The $GAIA$ models do not include the effects of dust, which can be a significant source of continuum opacity in late M-type and L-type objects (\citealt{Tsuji:1996p19056}; \citealt{Jones:1997p17642}; \citealt{Leinert:2000p19067}; \citealt{Allard:2001p14776}; \citealt{Pavlenko:2006p19075}).  

The AMES-Dusty grid (\citealt{Allard:2001p14776}) is a popular alternative set of low-temperature models which include the limiting effects of dust, but the sub-solar metallicity AMES-Dusty models only extend as cool as 2800 K.  Nevertheless, to gauge the differences between non-dusty and dusty models, we fit the solar-metallicity AMES-Dusty models ($\Delta$$T_\mathrm{eff}$ = 100 K; $\Delta$log $g$ = 0.5) to our $J$-band OSIRIS and 1.00-2.35 $\mu$m SpeX/prism spectra of HD 114762B.  The best-fitting AMES-dusty model to the OSIRIS spectrum is [2800/5.5] and to the SpeX spectrum is [2600/4.0].  These effective temperatures are similar to those of the best-fitting $GAIA$ models and so the use of non-dusty models does not appear to be the primary cause of the hotter temperatures.  However, it remains possible that both sets of models produce systematically hot effective temperatures.

An alternative explanation for the hotter effective temperatures is the metal-poor nature of HD 114762B compared to the solar-metallicity M dwarfs used in previous studies.  As we discuss in $\S$ 3.2.4, ultracool subdwarfs appear to be overluminous compared to field ultracool dwarfs of the same spectral class, possibly a result of a hotter effective temperature scale for ultracool subdwarfs.  A systematic study of ultracool subdwarf effective temperatures using other methods, like the infrared flux method, will provide insight into this discrepancy.

\subsubsection{The Luminosity of HD 114762B}

We calculated the luminosity of HD 114762B using its bolometric flux and the (revised) HIPPARCHOS parallax measurement  of HD 114762A (25.87 $\pm$ 0.76 mas, or 38.65$^{+1.17}_{-1.10}$ pc; \citealt{vanLeeuwen:2007p12454}; \citealt{Perryman:1997p534}).   The bolometric flux was computed twice using the original and shifted SpeX spectra to test whether shifting the spectrum significantly influenced the resulting luminosity.   To determine the bolometric flux of the unshifted data we first created an artificial spectrum by adding noise drawn from the measurement errors.  We then used the best-fitting model as a bolometric correction by attaching a short-wavelength segment (10$^{-3}$ $\mu$m $<$ $\lambda$  $<$ 1.00 $\mu$m) and a long-wavelength segment (2.35 $\mu$m $<$ $\lambda$ $<$ 10$^3$ $\mu$m) to the artificial spectrum.  To account for the uncertainty in the flux calibration we adjusted the spectrum by a scaling factor drawn from a Gaussian distribution with a mean $\overline{C}_\mathrm{fc}$ and error $\sigma_{\overline{C}_\mathrm{fc}}$.  The bolometric flux was then computed by integrating the spectrum.  This process was repeated 10,000 times, from which a mean bolometric flux and error were determined.  A similar process was carried out for the shifted spectrum.  Each artificial shifted spectrum consisted of $J$, $H$, and $K$ spectral regions to which noise was added.  In addition, short-wavelength (10$^{-3}$ $\mu$m $<$ $\lambda$ $<$ 1.10 $\mu$m) and long-wavelength (2.35 $\mu$m $<$ $\lambda$ $<$ 10$^3$ $\mu$m) segments from the best-fitting model were attached to the artificial data, as were model contributions between the  $J$ to $H$ and $H$ to $K$ filter bandpasses.  The spectrum was multiplicatively scaled by a Monte Carlo-generated flux-calibration scaling factor and the bolometric flux was computed.  This process was repeated 10,000 times.  

\begin{figure}
  \resizebox{3.5in}{!}{\includegraphics{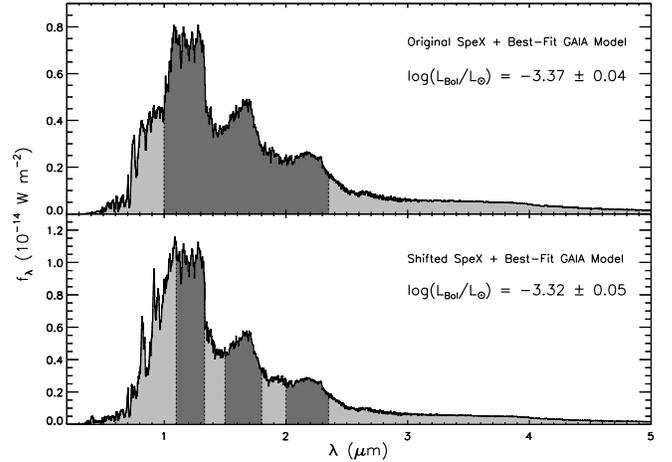}}
  \caption{Bolometric corrections for HD 114762B.  The top panel shows the original SpeX/prism spectrum (dark grey) and the best-fitting model used for the bolometric correction (light grey; [[M/H]/$T_\mathrm{eff}$/log $g$] = [--2.0/2500/1.0]). The bottom panel shows the SpeX spectral segments shifted to the published photometry (dark grey) and the best-fitting model for the correction ([+0.5/2600/3.0]).  The models have been smoothed to the resolution of the data. \label{f15}} 
\end{figure}

The uncertainty in the bolometric luminosity incorporates the error in the parallactic distance measurement ($\sigma_d$) and the error in the bolometric flux ($\sigma_{F_\mathrm{Bol}}$) in the following way:

\begin{equation}
\sigma_{L_\mathrm{Bol}} = 4 \pi d^2 F_\mathrm{Bol}   \sqrt{    \left(   \frac{ \sigma_{F_\mathrm{Bol}} }{F_\mathrm{Bol}}\right)^2    +  \left( \frac{2 \sigma_{d}}{d}   \right)^2          }.
\end{equation}

\noindent The resulting bolometric flux and luminosity from the original spectrum are 9.11 $\pm$ 0.83 $\times$ 10$^{-15}$ W m$^{-2}$ and log($L_\mathrm{Bol}/L_{\odot}$) = --3.37 $\pm$ 0.04, and for the shifted SpeX spectrum are 10.32 $\pm$ 0.95 $\times$ 10$^{-15}$ W m$^{-2}$ and log($L_\mathrm{Bol}/L_{\odot}$) = --3.32 $\pm$ 0.04 (Figure \ref{f15}; Table \ref{atm_mod}).

 We also investigated the effect of systematic errors due to the choice of model used in the bolometric correction.  We recalculated the bolometric luminosities of the original and shifted SpeX spectra using the second and third best-fitting synthetic spectra.  The change in log($L_\mathrm{Bol}$/$L_{\odot}$) from choosing different model atmospheres was less than 0.01 dex for the original SpeX spectrum and was near 0.02 dex for the shifted spectrum.  In addition, we computed the luminosities using the [--0.5/2600/5.5] $GAIA$ model.  This model was chosen because its parameters are the closest to those predicted by the evolutionary models (see $\S$ 3.3).  The change in luminosiy was $\sim$ 0.01 dex for the original spectrum and was $\sim$ 0.02 dex for the shifted spectrum.  These errors are uncorrelated for high signal-to-noise spectra, so we added the systematic errors to the statistical errors, in quadrature, to obtain our final uncertainty estimates in the luminosities: 0.04 and 0.05 dex for the original and shifted spectra, respectively. \\ \\ \\

\begin{deluxetable*}{cccccc}
\tabletypesize{\scriptsize}
\tablewidth{0pt}
\tablecolumns{6}
\tablecaption{Luminosity and Radius of HD 114762B  \label{atm_mod} }
\tablehead{
\colhead{}   &    \colhead{}   &    \colhead{$\overline{C}_k$ ($\times$ 10$^{-21}$)}   &    \colhead{$R$ ($R_{\odot}$)}   &    \colhead{$R$ ($R_{\odot}$)}   &    \colhead{}   \\
\colhead{Spectrum}  & \colhead{[[M/H]/$T_\mathrm{eff}$/log $g$]}         &   \colhead{\{ = $R^2$/$d^2$\}}   &    \colhead{ \{ = $d \sqrt{\overline{C}_k}$ \} }   &   \colhead{ \{  =  $\sqrt{L_\mathrm{Bol}/(4 \pi \sigma T_\mathrm{eff}^4)}$  \} } &   \colhead{log($L_\mathrm{Bol}/L_{\odot}$)\tablenotemark{a}}   
}
\startdata
SpeX  (1.00 - 2.35 $\mu$m)       &    [--2.0/2500/1.0]    &   4.6 $\pm$ 0.4    &  0.116 $\pm$ 0.006   & 0.111 $\pm$ 0.010 &  --3.37 $\pm$ 0.04     \\
Shifted SpeX ($J$/$H$/$K$)   &  [0.5/2600/3.0]       &     4.4 $\pm$ 0.5    &  0.113 $\pm$ 0.007  & 0.110 $\pm$ 0.010   &    --3.32 $\pm$ 0.05 \\
\enddata
\tablenotetext{a}{We adopt a value of 3.86$\times$10$^{26}$ W m$^{-2}$ for $L_{\odot}$. \\}
\end{deluxetable*}

\begin{deluxetable*}{lcccccc}
\tabletypesize{\scriptsize}
\tablewidth{0pt}
\tablecolumns{7}
\tablecaption{Luminosities and Bolometric Corrections of Ultracool Subdwarfs \label{lumsd}}
\tablehead{
\colhead{}  &  \colhead{}  &  \colhead{Best-Fitting}&  \colhead{}   &  \colhead{$BC_J$}   & \colhead{$BC_{K_\mathrm{S}}$}    & \colhead{Spectrum}  \\
        \colhead{Object}   & \colhead{SpT} &   \colhead{$GAIA$ Model\tablenotemark{a}}     &   \colhead{log($L_\mathrm{Bol}/L_{\odot}$)} & \colhead{(mag)}   & \colhead{(mag)}     & \colhead{Ref.}    
}
\startdata
 LHS 377                        &  sdM7      &  [0.5/3200/5.5]     &  --3.08 $\pm$ 0.04   & 1.97 $\pm$ 0.04       &  2.69 $\pm$ 0.04  & 1 \\
 LSR 2036+5059          &  sdM7.5   &  [--1.0/2900/5.0]  &  --3.02 $\pm$ 0.05  &  2.01 $\pm$ 0.04       &  2.69 $\pm$ 0.04  &  2  \\
 SSSPM 1013--2356    & sdM9.5   &  [--1.5/2600/5.0]   &  --3.41 $\pm$ 0.08  & 2.07 $\pm$ 0.05       &  2.30 $\pm$ 0.09  &  1  \\
 2MASS 1626+3945     &  sdL4       &  [--2.5/2200/3.0]  &  --3.71  $\pm$ 0.04  & 2.21 $\pm$ 0.04      &  2.18 $\pm$ 0.08 & 1 \\
\enddata
\tablenotetext{a}{For fits to the 0.80-2.35 $\mu$m spectral region; model parameters are [[M/H]/$T_\mathrm{eff}$ (K)/log $g$].}
\tablerefs{(1) \citet{Burgasser:2004p564}, (2) \citet{Burgasser:2004p574} }
\end{deluxetable*}

\subsubsection{The Luminosities of Other Ultracool Subdwarfs}

Using the same method described in \S3.2.3, we calculated the luminosities of four other ultracool subdwarfs with published parallaxes and whose spectra were available in the SpeX Prism Spectral Library: LHS 377 (sdM7; \citealt{Monet:1992p2839}, \citealt{Gizis:1997p80}), LSR 2036+5059 (sdM7.5; \citealt{Lepine:2003p4126}), SSSPM 1013--1356 (sdM9.5; \citealt{Scholz:2004p579}), and 2MASS 1626+3945 (sdL4; \citealt{Burgasser:2004p564}).  The spectra were flux calibrated using their 2MASS $J$-band magnitudes and the best-fitting $GAIA$ models were used to determine bolometric corrections.  Parallaxes for LSR 2036+5059, SSSPM 1013--1356, and 2MASS 1626+3945 are from \citet{Schilbach:2009p14779} and the parallax for LHS 377 is from \citet{Monet:1992p2839}.  The spectra were originally published by \citet{Burgasser:2004p564} and \citet{Burgasser:2004p574}. The best-fitting models used for the short- and long-wavelength corrections were [0.5/3200/5.5], [--1.0/2900/5.0], [--1.5/2600/5.0], and [--2.5/2200/3.0] for LHS 377, LSR 2036+5059, SSSPM 1013--1356, and 2MASS 1626+3945, respectively.  We added in quadrature the resulting luminosity measurement errors and an estimated systematic error of 0.02 dex to account for the choice of atmospheric model used for the bolometric correction.  The results are summarized in Table \ref{lumsd}.   The luminosity of LHS 377 has previously been calculated by \citet{Leggett:2000p560} using an observationally-derived $K$-band bolometric correction; the luminosity we obtained (--3.08 $\pm$ 0.04 dex) is consistent with their value of --3.11 dex. The luminosity of HD 114762B (--3.37 $\pm$ 0.04 dex) is similar to that of the sdM9.5 object SSSPM 1013--1356 (--3.41 $\pm$ 0.08 dex), which is consistent with the spectral type of d/sdM9 $\pm$ 1 that we assigned to HD 114762B.

\begin{figure}
  \resizebox{3.45in}{!}{\includegraphics{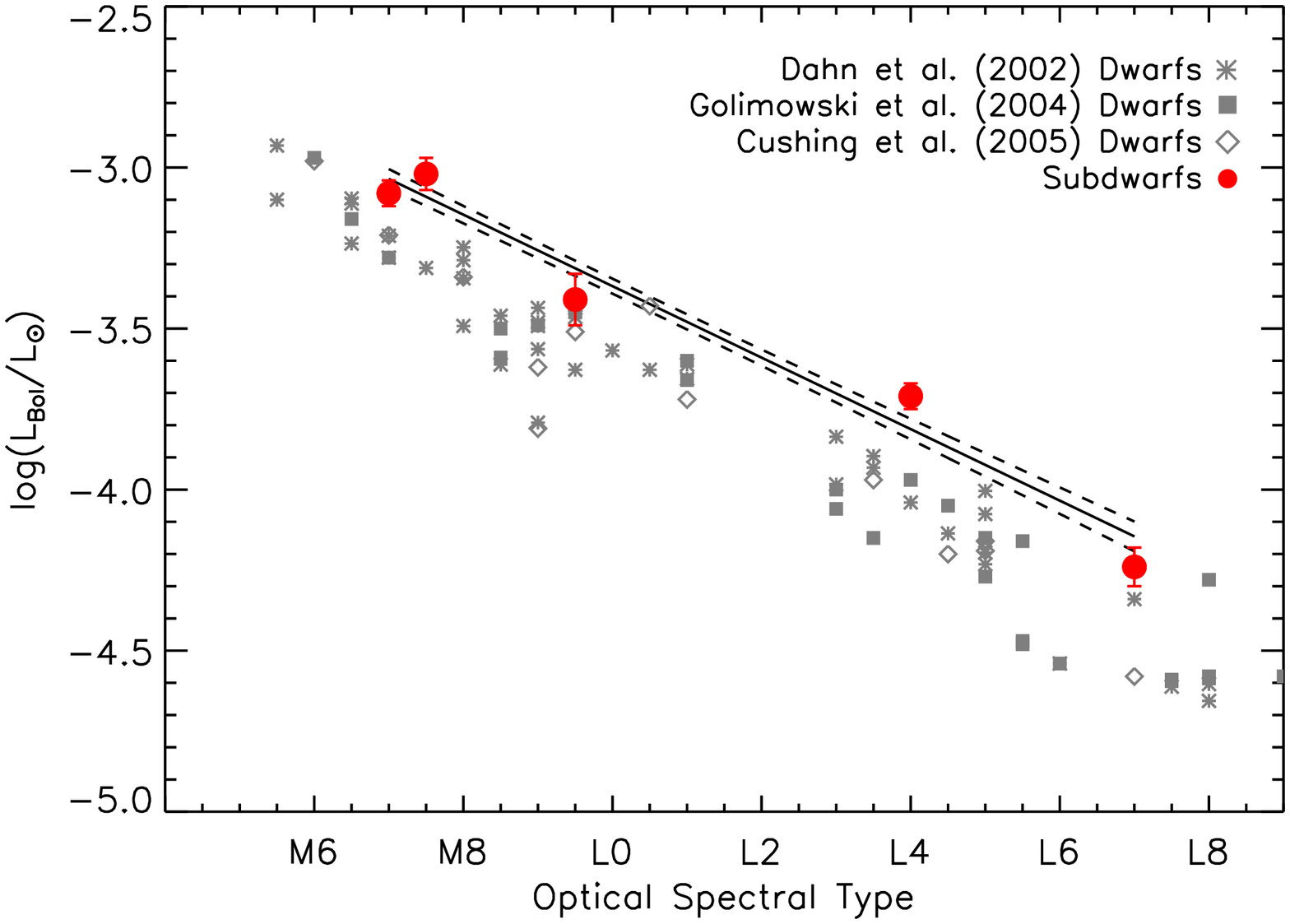}}
  \caption{Bolometric luminosity as a function of spectral type for ultracool dwarfs (gray) and ultracool subdwarfs (red).  The data for the dwarfs are from \citet{Dahn:2002p13692}, \citet{Golimowski:2004p15703}, and \citet{Cushing:2005p288}.  Objects that have since been discovered to be resolved close binaries have been removed from the samples.  The subdwarf data are from this work (Table \ref{lumsd}) and \citet[for 2MASS 05325346+8246465 (sdL7)]{Burgasser:2008p2494}.  The solid black line and the dashed black curves show the linear fit and $\pm$ 1 $\sigma$ uncertainty (Equation 9).  \label{f16}} 
\end{figure}

In Figure \ref{f16} we plot the bolometric luminosity as a function of optical spectral type for dwarfs (from \citealt{Dahn:2002p13692}, \citealt{Golimowski:2004p15703}, and \citealt{Cushing:2005p288}) and subdwarfs (from this work and \citealt{Burgasser:2008p2494}).  Objects that have since been resolved as close binaries have been removed from the samples.  For a given optical spectral type, the luminosities of ultracool subdwarfs appear to be slightly higher than those of dwarfs.  A linear fit to the ultracool subdwarf luminosities from spectral types M7 to L7 yields the following coefficients:

\begin{equation}
\log(L/L_{\odot}) = -2.258 - 0.111\times SpT, 
\end{equation}
\begin{displaymath}
Cov = \left(
\begin{array}{cc}
4.34 & -0.369 \\
-0.369 & 0.0357
\end{array}\right) \times 10^{-3},
\end{displaymath}

\noindent where $SpT$ is the ultracool subdwarf numerical spectral type beginning at M0 = 0 and increasing by 1 for every spectral subclass (e.g., L5 = 15).  $Cov$ is the covariant matrix of the fit.

There are several explanations for the apparent overluminosity of ultracool subdwarfs in the luminosity-spectral type diagram.  Unresolved binarity is one possibility, but it is unlikely that the few late-type subdwarfs sampled thus far all happen to be unresolved binaries.    An alternative explanation is that the effective temperatures and/or radii of ultracool subdwarfs are different than those of normal dwarfs in the same spectral subclass.  Ultracool subdwarfs appear to be $\sim$ 0.2 dex above the luminosity-spectral type trend of ultracool dwarfs.  If the higher luminosity were due to a larger radius alone then ultracool subdwarfs would need to have radii that were $\sim$ 26\% larger than those of dwarfs.  If, on the other hand, the increased luminosity were caused by a difference in effective temperature, then ultracool subdwarfs would need effective temperatures that were $\sim$ 12\% hotter than dwarfs from the same spectral subclass.  This translates into temperature differences of $\sim$ 150-250 K for objects throughout the late-L to late-M spectral sequence.   \citet{Burgasser:2006p2516} found additional evidence for different effective temperature scales for ultracool dwarfs and subdwarfs by comparing their atmospheric model-derived effective temperatures to their spectral types.  They found that extreme M-type subdwarfs appear to have higher effective temperatures by several hundred Kelvin compared to normal M dwarfs, which they suggest might originate from differing spectral classification schemes.  Unequal effective temperature scales for objects with the same spectral type but different luminosity class is not new: M dwarfs and M giants, for example,  to exhibit this trait (see, e.g., \citealt{diBenedetto:1993p17643}). 

\subsubsection{Bolometric Corrections for Ultracool Subdwarfs}

Using our computed bolometric fluxes, we can derive bolometric corrections for ultracool subdwarfs.  The $J$-band bolometric correction is defined as $BC_J$ $\equiv$ $M_\mathrm{Bol}$ -- $M_J$ = $m_\mathrm{Bol}$ -- $J$, where $M_\mathrm{Bol}$ is the absolute bolometric magnitude, $M_J$ is the $J$-band absolute magnitude, $m_\mathrm{Bol}$ is the apparent bolometric magnitude, and $J$ is the $J$-band apparent magnitude.  Following \citet[Appendix D]{Bessell:1998p18664}, we define $M_{\mathrm{Bol},{\odot}}$ to be 4.74 and we assume $L_{\mathrm{Bol},{\odot}}$ = 3.86 $\times$ 10$^{26}$ W, in which case the $J$-band bolometric correction becomes $BC_J = -2.5 \log(F_\mathrm{Bol}) - 18.988 - J$, where $F_\mathrm{Bol}$ is the bolometric flux in W.

\begin{figure}
  \resizebox{3.45in}{!}{\includegraphics{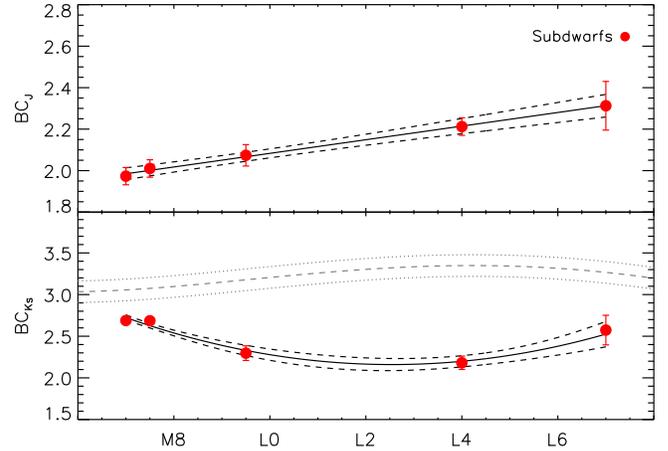}}
  \caption{$J$- and $K_\mathrm{S}$-band bolometric correction as a function of spectral type for ultracool subdwarfs.  The gray dashed curve in the bottom panel shows the $K$-band bolometric correction (not $K_\mathrm{S}$) for ultracool dwarfs derived by \citet{Golimowski:2004p15703}.  The dotted gray lines represent the typical rms scatter of ultracool dwarfs (0.13 mags) about the best-fitting polynomial relation.  Ultracool subdwarf names and their individual corrections are tabulated in Table \ref{lumsd}.  The coefficients and covariance matrices of the fits are given in Equations 10 and 11.  These relations are only applicable for mildly metal-poor ``sd''-type subdwarfs.  \label{f17}} 
\end{figure}

$J$- and $K_\mathrm{S}$-band bolometric corrections for ultracool subdwarfs are  plotted in Figure \ref{f17} and are tabulated in Table \ref{lumsd}.  We use the values from \citet{Burgasser:2008p2494} for 2MASS 05325346+8246465 (sdL7) to derive bolometric corrections for that object.  A linear fit to the $J$-band bolometric corrections yield the following relation: 

\begin{equation}
BC_J = 1.75 + 0.0328 \times SpT,
\end{equation}
\begin{displaymath}
Cov = \left(
\begin{array}{cc}
4.95 & -0.461  \\
-0.461 & 0.0474  
\end{array}\right) \times 10^{-3},
\end{displaymath}

\noindent where $SpT$ is the numerical spectral type beginning at M7 = 7 and increasing by 1 for each spectral subclass through L7.  The rms scatter about the fit is 0.007 mags.  Similarly, a quadratic fit to the $K_\mathrm{S}$ bolometric correction gives 

\begin{equation}
BC_{K_\mathrm{S}} = 5.05 - 0.461 \times SpT + 0.0184 \times SpT^2,
\end{equation}
\begin{displaymath}
Cov = \left(
\begin{array}{ccc}
292 & -59.1 & 2.66 \\
-59.1 & 12.0 & -0.545 \\
2.66 & -0.545 & 0.0249 
\end{array}\right) \times 10^{-3}.
\end{displaymath}

\noindent The rms scatter about the quadratic fit is 0.04 mags.  We note that these relations are only applicable for moderately metal-poor ultracool subdwarfs and should not be applied for ``usd''- or ``esd''-type objects.

\begin{figure*}
  \plotone{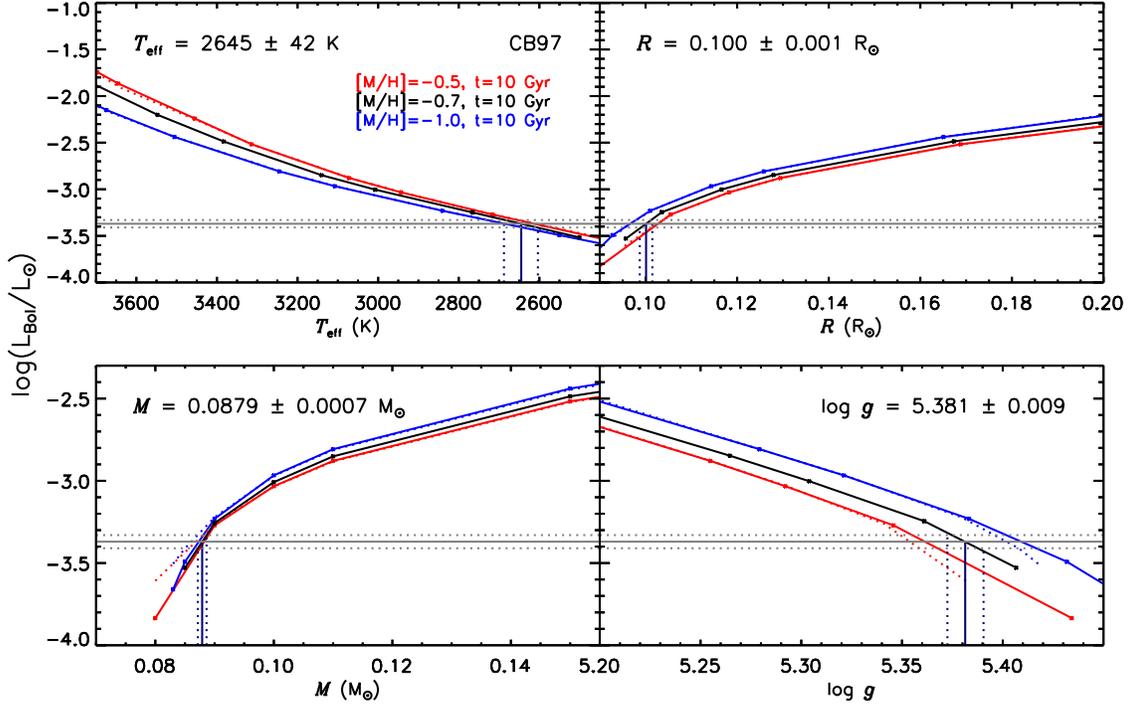}
  \caption{Predictions from the CB97 evolutionary models based on the bolometric luminosity of HD 114762B (log($L_\mathrm{Bol}/L_{\odot}$) = --3.37 $\pm$ 0.04, solid and dotted gray lines) and the metallicity of HD 114762A ([Fe/H] = --0.71 $\pm$ 0.07). The [M/H] = --0.5 (red) and --1.0 (blue) metallicity tracks are shown for 1 Gyr (dotted lines) and 10 Gyr (solid lines).  An interpolated --0.7 dex 10 Gyr metallicity track is shown in black.   Mean values and errors (solid and dotted navy lines) for the effective temperature,  radius, mass, and surface gravity are derived by performing a Monte Carlo simulation of luminosity and metallicity values.  \label{f18}} 
\end{figure*}

\subsubsection{The Radius of HD 114762B}

The radius of HD 114762B can be calculated in two ways using the model atmospheres.  One way is to use the scaling factor $\overline{C}_k$ which scales the emergent flux density of  the best-fitting model to the flux-calibrated spectra.  The radius $R$ and the error $\sigma_R$  can be computed using the following relations:

\begin{equation}
R = d \sqrt{\overline{C}_k },
\end{equation}

\begin{equation}
\sigma_R = R   \sqrt{  \left(  \frac{ \sigma_{\overline{C}_k}}{2 \overline{C}_k} \right)^2      +    \left(  \frac{ \sigma_{d} }{d}     \right)^2     },
\end{equation}
 
\noindent where $d$ and $\sigma_d$ are the distance and uncertainty in the distance to the object, and $\overline{C}_k$ and $\sigma_{\overline{C}_k}$ are the model scaling factor and its error.  The radii computed using this method are 0.116 $\pm$ 0.006 $R_{\odot}$ and 0.113 $\pm$ 0.007 $R_{\odot}$ for the original and shifted SpeX spectra, respectively, where $R_{\odot}$ = 6.960 $\times$ 10$^{10}$ cm.

The radius may also be computed using the bolometric luminosity and the effective temperature derived from the best-fitting atmospheric model:

\begin{equation}
R = \left( \frac{L_\mathrm{Bol}}{4 \pi \sigma T_\mathrm{eff}^4 } \right)^{1/2}
\end{equation}

 \begin{equation}
 \sigma_R = R \sqrt{ \left( \frac{\sigma_{L_\mathrm{Bol}}}{2 L_\mathrm{Bol}} \right)^2 + \left( \frac{2 \sigma_{T_\mathrm{eff}}}{T_\mathrm{eff}} \right)^2  }
 \end{equation}

 In this work we do not estimate the errors of the fundamental parameters derived from fitting atmospheric models; instead we merely report the best few models as a triplet of [[M/H], $T_\mathrm{eff}$, log $g$].  To compute the uncertainty of the radius we estimate $\sigma_{T_\mathrm{eff}}$ to be $\sim$ 100 K based on the narrow range of resulting temperatures from the best-fitting models.  This gives radii of 0.111 $\pm$ 0.010 $R_{\odot}$ and 0.110 $\pm$ 0.010 $R_{\odot}$ for the original and shifted SpeX spectra, respectively.  These values are consistent with the radii derived using the model scaling factor.

\subsection{Comparing Physical Parameters From Atmospheric and Evolutionary Models}

Evolutionary models predict the change in stellar luminosity, effective temperature, and radius as a function of mass, age, and metallicity.  When testing evolutionary models, the interpretation of discrepancies between observations and model predictions must be made with care.  Evolutionary models depend on atmospheric models to account for the way that radiative flux escapes from a central luminosity source, so the choice of atmospheric models can influence the predictions of the evolutionary models.  It may therefore be difficult to untangle inadequate atmospheric models from missing or incorrect physics in the evolutionary models.  

These models can be unraveled to some extent by studying their predictions in various diagrams.  The fundamental physical parameters predicted by the evolutionary models ($L_\mathrm{Bol}$, $T_\mathrm{eff}$, and $R$) are more dependent on the input physics than on the atmospheric models used (\citealt{Chabrier:2000p161}).  However, converting $L$ and $T_\mathrm{eff}$ into observable quantities (magnitudes and colors) relies on the emergent spectral energy distributions, so the predictive accuracy of color-magnitude diagrams reflects the accuracy of the atmospheric models more so than do $L$-$T_\mathrm{eff}$ HR diagrams.  This dependency on correct atmospheric models is especially strong for low-mass stars and brown dwarfs, for which the inclusion of clouds in the models is required to even approximately match the observed trends of late-M, L, and T dwarfs in the near-infrared color-absolute magnitude diagrams (\citealt{Allard:2001p14776}; \citealt{Tsuji:2003p15622}; \citealt{Burrows:2006p7009}; \citealt{Saumon:2008p14070}).  In this study we  use the fundamental parameters from by the \citet[CB97]{Chabrier:1997p2767} models and the color-magnitude relations from the \citet[BCAH97]{Baraffe:1997p582} and \citet[BCAH98]{Baraffe:1998p160} models, all of which incorporate the same (PHOENIX/$NextGen$) model atmospheres and are based on the same input physics.  These models are widely used throughout the literature so it is crucial to test them at low masses and metallicities, a task that we perform here for the first time.

\begin{figure*}
  \plotone{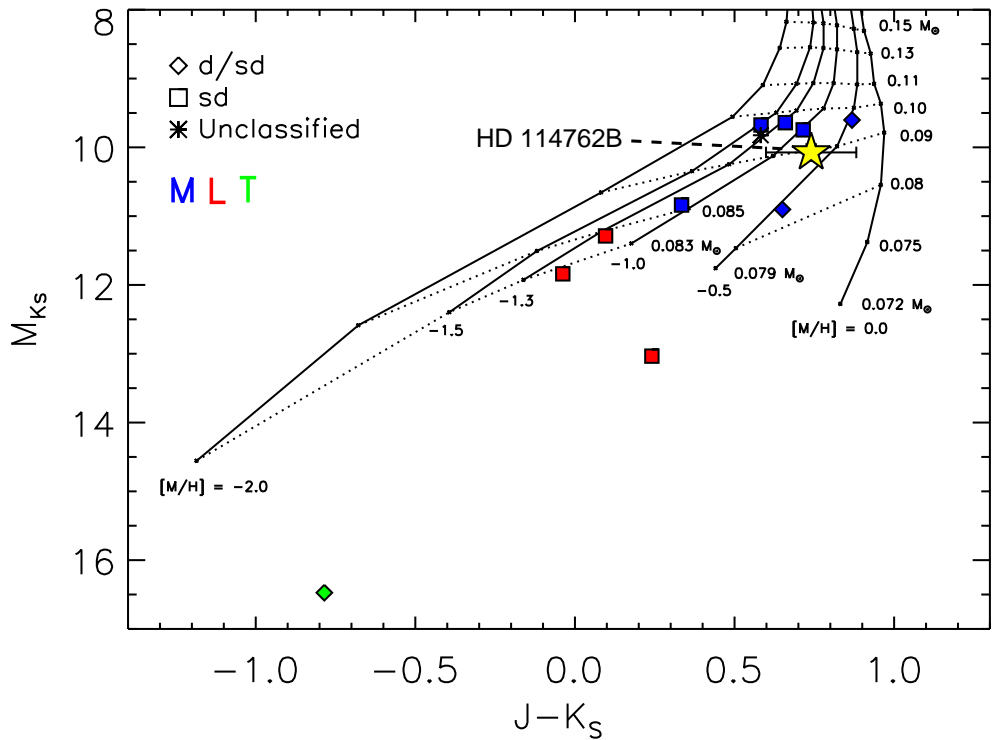}
  \caption{$M_{K_\mathrm{S}}$ vs $J$ -- $K_\mathrm{S}$ color-absolute magnitude diagram showing the positions of ultracool subdwarfs with parallaxes. The [M/H] = \{0.0, --0.5\} evolutionary tracks are from BCAH98 and the [M/H] = \{--1.0, --1.3, --1.5, --2.0\} tracks are from BCAH97 (10 Gyr models for both).  The evolutionary  tracks were converted from the CIT photometric system to the 2MASS system using the relations in \citet{Carpenter:2001p159}.  Parallaxes are from \citet{Schilbach:2009p14779} and \citet{Monet:1992p2839}.  See Figure 1 of \citet{Schilbach:2009p14779} for the identification of the other ultracool subdwarfs in this diagram.  The position of HD 114762B in this diagram is shown as a yellow star.  Its predicted metallicity from the color-magnitude diagram is consistent with the value of --0.70 dex for the primary star HD 114762A.  \label{f19}} 
\end{figure*}

Using the metallicity and the now well-measured luminosity of HD 114762B, we compared the effective temperatures, radii, and surface gravities from our atmospheric model fitting to the values predicted by the evolutionary models to test for consistency.\footnote{While masses can be inferred from the atmospheric models using the radius and surface gravity, the results are not practical for comparison because the models are coarsely gridded ($\Delta$log $g$ = 0.5).}  Mutually consistent predictions do not prove that the models are correct (both may predict the wrong values), but inconsistent results indicate that at least one set of models needs improvement.

\begin{deluxetable}{cccc}
\tabletypesize{\scriptsize}
\tablewidth{0pt}
\tablecolumns{4}
\tablecaption{Results From CB97 Evolutionary Models  for HD 114762B \label{ev_mod}}
\tablehead{
        \colhead{$T_\mathrm{eff}$ (K)}   &    \colhead{$R$ ($R_{\odot}$)}     &   \colhead{$M$ ($M_{\odot}$)}     &  \colhead{log $g$}      
}
\startdata
 2645 $\pm$ 42   &  0.100 $\pm$ 0.001   &  0.0879 $\pm$ 0.0007   &  5.381  $\pm$ 0.009    \\
\enddata
\tablecomments{Errors are derived in a Monte Carlo fashion using the bolometric luminosity from the original SpeX data (log($L_\mathrm{Bol}/L_{\odot}$) = --3.37 $\pm$ 0.04) and a metallicity of --0.71 $\pm$ 0.07 dex.}
\end{deluxetable}

There are a large number of age estimates in the literature for the primary star HD 114762A, the majority of which are based on evolutionary model isochrones or chromospheric activity levels.  Most estimates are greater than 10 Gyr (Table \ref{metallicity}).  For the low mass evolutionary tracks that are relevant to HD 114762B, there is little difference between the 1 Gyr and 10 Gyr isochrones (Figure \ref{f18}).  We chose the 10 Gyr evolutionary tracks for our analysis.  The properties of HD 114762B were derived in a Monte Carlo fashion by first randomly choosing a luminosity from a Gaussian distribution with mean values of --3.37 dex and a standard deviation of 0.04 dex.  Similarly, a metallicity was chosen from a Gaussian distribution with a mean value of --0.71 dex and a standard deviation of 0.07 dex.  We created a new metallicity track at the chosen value by interpolating the evolutionary models.  Then using the luminosity and the new metallicity track, we determined a temperature, radius, mass, and surface gravity for that trial.  This process was repeated 10,000 times and the resulting distributions are summarized in Table \ref{ev_mod}.  The CB97 models yielded the following results for log($L_\mathrm{Bol}/L_{\odot}$) = --3.37 $\pm$ 0.04 and [M/H] = --0.71 $\pm$ 0.07: $T_\mathrm{eff}$ = 2645 $\pm$ 42 K, $R$ = 0.100 $\pm$ 0.001 $R_{\odot}$, $M$ = 0.0879 $\pm$ 0.0007 $M_{\odot}$, and log $g$ = 5.381 $\pm$ 0.009.

The $GAIA$ atmospheric models and the CB97 evolutionary models generally predicted consistent effective temperatures (between 2500 and 2700 K for the former and 2645 K for the latter) and radii, but the inferred surface gravities were largely inconsistent.  The exceptions were the fits to the OSIRIS data.  Fitting the entire OSIRIS spectrum with the $GAIA$ models resulted in a temperature of 2800 K (or $\sim$ 150 K above that of the CB97 evolutionary models) and a surface gravity of 5.5 dex (similar to 5.38 dex from the CB97 evolutionary models).  The fit to the $J$-band absorption lines yielded $T_\mathrm{eff}$ = 2600 K and log $g$ = 5.0.  \citet{Jao:2008p6434} used the radii and masses of low-mass stars from  \citet{LopezMorales:2007p4016} to construct a mass-gravity relationship, which exhibited an increasing trend of surface gravity with lower mass.  For a mass of $\sim$ 0.09 $M_{\odot}$, we can therefore expect HD 114762B to have a surface gravity above $\sim$ 5.0 dex.  Most of the surface gravities from fits to the SpeX data were unrealistically low (log $g$ $\lesssim$ 4.0) compared to the empirical trends and compared to the evolutionary models.  Finally, we note that the radius inferred from the atmospheric models ($\sim$ 0.11 $R_{\odot}$) is statistically slightly larger than the value predicted by the evolutionary models ($\sim$ 0.10 $R_{\odot}$).  

The color-magnitude diagram (CMD) provides another test of the atmospheric models.  In Figure \ref{f19} we plot the position of HD 114762B with respect to the BCAH97 and BCAH98 evolutionary models in the M$_{K_\mathrm{S}}$ vs $J$ -- $K_\mathrm{S}$ CMD.  The [M/H] = \{0.0, --0.5\} metallicity tracks are from BCAH98 while the [M/H] = \{--1.0, --1.3, --1.5, --2.0\} tracks are from BCAH97.  These evolutionary models make use of an earlier version of the PHOENIX code (the \emph{NextGen} models) than the $GAIA$ grid that we studied in this work.  The location of HD 114762B in this diagram is comfortably between the --0.5 and --1.0 dex metallicity tracks, which is in precise agreement with the metallicity of --0.70 dex of the primary star.  Previously we showed that the synthetic spectra from the atmospheric models did a poor job of reproducing the correct physical parameters of HD 114762B from the best-fitting models.  Nevertheless, the color-absolute magnitude diagram, which crudely measures the synthetic spectral energy distributions of the models, appears to well-match the observations.  This suggests that the shape of the models at lower metallicities is accurate, but that the detailed treatment of opacity sources in the near-infrared still needs improvement.

To obtain a sense of the relative metallicities of other metal-poor low-mass objects, we also plot all currently known ultracool subdwarfs with parallax measurements in the M$_{K_\mathrm{S}}$ vs $J$ -- $K_\mathrm{S}$ CMD in Figure \ref{f19}.  Parallaxes are from \citet{Schilbach:2009p14779} and \citet[for LHS 377]{Monet:1992p2839}.  Spectroscopically-classified mild subdwarfs fall in the expected mild-metallicity regions of the BCAH97/98 evolutionary tracks, and objects classified as subdwarfs are located between the [M/H] = --1.0 to  --2.0 metallicity tracks.  The M, L, and T spectral types appear to follow an increasing absolute magnitude trend with later spectral type, an expected feature for progressively later-type and lower-luminosity objects.  The locations of ultracool subdwarfs in the M$_{K_\mathrm{S}}$ vs $J$ -- $K_\mathrm{S}$ CMD therefore supports the current extension of the optical spectral classification scheme to lower metallicities.  Nevertheless, the development of a rigorous spectral classification scheme will be needed with the discovery of large numbers of ultracool subdwarfs.  

Based on the correct metallicity prediction for HD 114762B, we suspect that the accuracy of the BCAH97/98 evolutionary models is greater for ultracool subdwarfs than for their solar-metallicity counterparts, for which the inclusion of clouds in the atmospheric models is necessary to reproduce their observed $J$ -- $K_\mathrm{S}$ colors.  In the low-mass metal-poor regime, continuum CIA H$_2$ is the dominant opacity source in the near-infrared, so dust-free model atmospheres like $NextGen$ probably do a reasonable job of reproducing the color-magnitude relations of ultracool subdwarfs.  The discovery of additional benchmark companions over a large range of metallicities will be necessary for this claim to be properly verified. \\ \\

\begin{figure}
  \resizebox{3.45in}{!}{\includegraphics{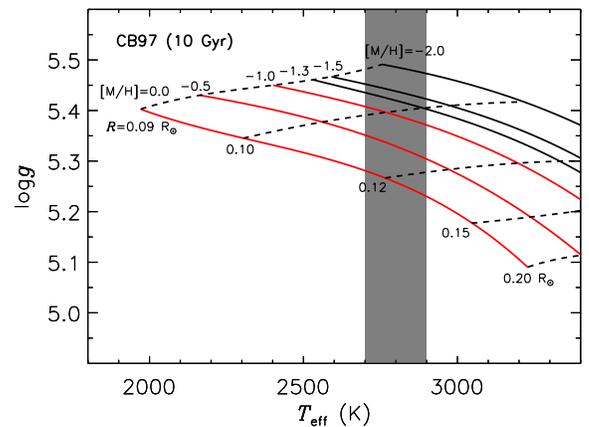}}
  \caption{Surface gravity as a function of effective temperature for different values of radius and metallicity from the 10 Gyr evolutionary models of \citet{Chabrier:1997p2767}.  The models are interpolated over radius and metallicity.  The gray shaded region shows the range of $T_\mathrm{eff}$ from fitting synthetic spectra to the OSIRIS data (2800 $\pm$ 100 K).  The red tracks show the metallicity range from the atmospheric model fitting (--0.5 $\pm$ 0.5 dex).  The evolutionary models constrain the radius, which, combined with the model scaling factor $\overline{C}_k$ (= $R^2/d^2$), can be used to estimate the distance to HD 114762B without knowing the parallax or the bolometric flux and assuming the object is not a binary.    \label{f20}} 
\end{figure}

\subsection{Spectroscopic Parallaxes}

\subsubsection{Spectroscopic Parallax to HD 114762B}

Since the factor $\overline{C}_k$ used to scale the model atmospheres to the flux calibrated spectra is equal to $R^2/d^2$, in principle accurate radius values could be used to derive the distance without knowing the parallax, assuming the object is not a binary. Evolutionary models can be used to constrain a radius if an age, effective temperature, and metallicity are known.
We have shown that, for HD 114762B, the most reliable technique for estimating the metallicity, surface gravity, and effective temperature from fitting synthetic spectra is to use the the medium-resolution $J$-band region.  Knowing \emph{a posteriori} that this fitting technique is acceptable, we can use the resulting effective temperature and metallicity to derive a radius from the evolutionary models and then a distance from $\overline{C}_k$ (equal to $R^2/d^2$).  This method is particularly well suited for observations of ultracool subdwarfs, which have ages that can be constrained from their kinematic properties.  We refer to the distances estimated using this method as ``spectroscopic parallaxes.''

The best-fitting model to our OSIRIS spectrum is [--0.5/2800/5.5] (Table \ref{tabfit}), where, for this analysis, we assume errors of 0.5 dex for the metallicity, 100 K for the effective temperature, and 0.5 dex for the surface gravity.  For ultracool objects, the evolutionary models provide tighter constraints on the surface gravity than the atmospheric models do, so the surface gravity is ignored for our estimate of the radius.  The evolutionary models predict the radius to be between $\sim$ 0.10 and 0.13 $R_{\odot}$ for an effective temperature of 2800 $\pm$ 100 K, a metallicity of --0.5 $\pm$ 0.5 dex, and an age of 10 Gyr, as shown in Figure \ref{f20}.  We obtained a value of 3.82 $\pm$ 0.35 $\times$ 10$^{-21}$ for $\overline{C}_k$ based on the best-fitting model to the flux-calibrated OSIRIS data ($\S$3.2.1).  For a radius of 0.115 $\pm$ 0.015 $R_{\odot}$, the distance estimate is 42 $\pm$ 6 pc, where the errors have been propagated as in $\S$3.2.5.  The parallactic distance to HD 114762A is 38.7 $\pm$ 1.1 pc.  The spectroscopic parallax method using the OSIRIS data therefore predicts the distance to HD 114762B to within 10\% of the parallactic distance. 

 We note that this spectroscopic parallax technique is not strongly sensitive to effective temperatures cooler than $\sim$ 3000 K.  Although the distance is directly proportional to the radius, the tracks of constant radius tend to spread out at in log $g$-$T_\mathrm{eff}$ space so that a small change in temperature produces a small change in radius in this low-temperature regime (Figure \ref{f20}).  At hotter temperatures, however, a small change in temperature produces a large change in radius; the uncertainty in radius is therefore impractically large for $T_\mathrm{eff}$ $\gtrsim$ 3000 K.  If we use a temperature of 2300 $\pm$ 100 K instead of 2800 K $\pm$ 100 K for HD 114762B (to better match temperatures of late M dwarfs in the literature; see $\S$ 3.2.2), the resulting (extrapolated) radius is $\sim$ 0.094 $\pm$ 0.009 (for a metallicity of --0.5 $\pm$ 0.5 dex), and the distance estimate using the spectroscopic parallax technique for this cooler temperature is 34 $\pm$ 4 pc.  This result is still consistent with the parallactic distance of HD 114762A at $<$ 2 $\sigma$.

\subsubsection{Spectroscopic Parallaxes to Other Ultracool Subdwarfs}

We applied this same spectroscopic parallax technique to four ultracool subdwarfs with parallaxes previously discussed in $\S$3.2.4 (LHS 377, LSR 2036+5059, SSSPM 1013--1356, and 2MASS 1626+3945).  As we have shown, the only reliable physical parameter from fitting atmospheric models to the SpeX spectra is the effective temperature.  Nevertheless, it is still possible to obtain an estimate of the radius using the effective temperature by assuming metallicities less than solar but greater than --2.0 dex (which encompasses the moderately metal-poor ``sd'' class of ultracool subdwarfs), ages of $\sim$10 Gyr, and errors of $\pm$ 100 K for the effective temperatures.  Using the effective temperatures from the atmospheric model fits (Table \ref{lumsd}), we obtained radii estimates of 0.16 $\pm$ 0.07 $R_{\odot}$,  0.12 $\pm$ 0.03 $R_{\odot}$,  0.100 $\pm$ 0.015 $R_{\odot}$, and  0.090 $\pm$ 0.010 $R_{\odot}$ for LHS 377, LSR 2036+5059, SSSPM 1013--1356, and 2MASS 1626+3945, respectively.     The resulting distance estimates are listed in Table \ref{specpartab} along with the parallactic distances.  The estimates from spectroscopic parallaxes are surprisingly accurate, although less so at higher effective temperatures ($\gtrsim$ 3000 K) where the models spread out in surface gravity space and poorly constrain the radii values (Figure \ref{f20}). The precision increases at lower temperatures for the same reason\footnote{We note that \citet{Leggett:2000p560} obtained an effective temperature of 2900 $\pm$ 100 K for LHS 377 using $T_\mathrm{eff}$-($V$--$I$)-[M/H] relations from evolutionary models.  A visual inspection of the best-fitting $GAIA$ model shows poor agreement between the LHS 377 spectrum and the model, so our value of 3200 K from the best-fitting $GAIA$ model may be overestimated.  A reduced $T_\mathrm{eff}$ for LHS 377 would lower the radius obtained from the evolutionary models, which would place the estimated distance closer to the actual value.  This may suggest that $GAIA$ model fits to earlier-type ultracool subdwarfs are not reliable, although we have shown that consistent values were obtained for objects near d/sdM8-d/sdM9 types ($\S$3.2.2).}.

Our spectroscopic parallax technique can be compared to the conventional photometric distance estimates which rely on empirical relations between spectral type and absolute magnitude.  Polynomial fits to spectral type-absolute magnitude data for late-type objects typically produce rms errors of $\sim$ 0.3-0.4 mag (\citealt{Tinney:2003p14852}; \citealt{Liu:2006p14530}), which translate into distance errors of $\sim$ 15-20\%.  In comparison, our spectroscopic parallax method produces distances that are about twice as accurate ($\sim$ 10\% agreement with published parallaxes), although this method has only been tested on a small sample of subdwarfs.  Extending this approach to a broader range of objects and wavelength ranges may be fruitful.

\begin{deluxetable}{lcccc}
\tabletypesize{\scriptsize}
\tablewidth{0pt}
\tablecolumns{5}
\tablecaption{Spectroscopic Parallaxes for Ultracool Subdwarfs \label{specpartab}}
\tablehead{
\colhead{}                        &  \colhead{$\overline{C}_k$ ($\times$ 10$^{-21}$) }   &  \colhead{$d_\mathrm{est}$\tablenotemark{a}}   &  \colhead{$d_{\pi}$\tablenotemark{b}}  & \colhead{$\pi$} \\
        \colhead{Object}    &  \colhead{ \{ = $R^2/d^2$ \} }                                            &  \colhead{(pc)}                                                           &  \colhead{(pc) }                                         &   \colhead{Ref.}    
}
\startdata
 LHS 377                        &    3.6 $\pm$  0.1 &  60 $\pm$ 30   & 35 $\pm$ 1   & 1 \\
 LSR 2036+5059          &    3.19 $\pm$ 0.09  & 50 $\pm$ 10   & 46 $\pm$  3  &  2  \\
 SSSPM 1013--2356    &   1.98 $\pm$ 0.06  & 51 $\pm$ \ 8    & 49 $\pm$ 5   &  2 \\
 2MASS 1626+3945     &   4.1 $\pm$ 0.1 & 32 $\pm$ \ 4     & 34 $\pm$ 1     & 2 \\
\enddata
\tablenotetext{a}{Spectroscopic parallax.  See $\S$3.4.}
\tablenotetext{b}{Parallactic distance.  Upper and lower limits on the distance errors are averaged for clarity.}
\tablerefs{(1) \citet{Monet:1992p2839}, (2) \citet{Schilbach:2009p14779}}
\end{deluxetable}

\section{Summary and Conclusion}

 We have presented a near-infrared spectroscopic analysis of HD 114762B, the first benchmark ultracool subdwarf companion.   The metallicity of HD 114762B is inferred from the primary star, which has a mean metallicity of [Fe/H] = --0.71 based on independent estimates from the literature.  The spectral characteristics of HD 114762B include suppressed $H$ and $K$-band continuum as well as deeper 1.4 and 1.9 $\mu$m H$_2$O bands compared to solar-metallicity M dwarfs.  The SpeX 1.00--2.35 $\mu$m spectrum does not, however, exhibit the extreme CIA H$_2$ which is characteristic of the ``sd'' class of ultracool subdwarfs.  In this respect, HD 114762B is more similar to mild subdwarfs, which are believed to have an intermediate metallicity between dwarfs and subdwarfs.  We assign it a spectral type of d/sdM9 $\pm$ 1 based on its resemblance to 2MASS 0041+35 (d/sdM9) from 1.00 to 2.35 $\mu$m.  A spectral type later than sdM7.5 can also be inferred  from the strength of the \ion{K}{1} lines and the strength of the 1.4 $\mu$m H$_2$O band compared to M-type subdwarf spectra.  The measured luminosity places HD 114762B just above the hydrogen-burning minimum mass based on the evolutionary models of CB97.
 
We use this unique  benchmark object to test low-mass metal-poor atmospheric and evolutionary models.  The best-fitting PHOENIX/$GAIA$ synthetic spectrum to our OSIRIS $J$-band data has physical parameters that are consistent with both the metallicity of the primary star and the high surface gravity of late-type objects.  Fits to the SpeX 1.00--2.35 $\mu$m spectrum, however, yield physical parameters that are inconsistent with these values.  We also fit the models to low-resolution spectra of other late-type mild ultracool subdwarfs with similar spectral types to HD 114762B.  The best-fitting models' surface gravity and metallicity are mutually inconsistent among the four objects and were likewise inconsistent for fits to different regions of the same spectrum.  Fits to HD 114762B, however,  yield consistent effective temperatures between 2500-2800 K.  We conclude that the metallicities and surface gravities derived from fitting atmospheric models to low-resolution near-infrared spectra of ultracool subdwarfs are not trustworthy, but that the effective temperatures are probably more reliable.

Problems with atmospheric models at low masses and low metallicities have already been noted in the literature; only moderately good matches have resulted from fits to the red-optical spectral regions of ultracool subdwarfs, even when applied over relatively short wavelength ranges (\citealt{Schweitzer:1999p563}; \citealt{Lepine:2004p559}; \citealt{Burgasser:2007p575}).   These same studies held the  model surface gravity fixed at log $g$ = 5.0 or 5.5 while only allowing the temperature and metallicity to vary.  The models should, however, be able to predict all three parameters with no \emph{a priori} assumption about any particular one.  In addition, the models should be able to predict roughly the same parameters for objects of the same spectral type as well as for different spectral regions of the same object.  We have shown that neither appears to be true for the ultracool subdwarfs we studied, making the discovery of benchmark metallicity calibrators like HD 114762B ever more important.  

Our results suggest that the best method to determine the metallicity of field ultracool subdwarfs may be to use the medium-resolution $J$-band region.  The 1.18-1.35 $\mu$m range encompasses metal-sensitive atomic absorption lines as well as the metal-sensitive 1.4 $\mu$m steam band.  An  interesting line that may be important for very metal-poor ultracool subdwarfs is the \ion{Fe}{1} 1.887 $\mu$m feature, which the $GAIA$ models predict to be essentially independent of temperature and gravity, but which is strongly dependent on metallicity for [M/H] $\lesssim$ --1.0.  We found that fitting the individual absorption lines rather than the entire 1.18-1.35 $\mu$m spectrum of HD 114762B produced temperatures and gravities that were more consistent with the values derived from the CB97 evolutionary models.  Nevertheless, fitting the entire $J$-band spectrum and the individual absorption features yielded metallicities consistent with the primary star and surface gravities consistent with empirical values of late-type objects.  

The location of HD 114762B in the BCAH97/98 M$_{K_\mathrm{S}}$ vs $J$ -- $K_\mathrm{S}$ color-absolute magnitude diagram is consistent with the metallicity of the primary star.   The metallicity and surface gravity inferred from $GAIA$ fits to low-resolution near-infrared spectra are not trustworthy, but the results from the color-magnitude diagram suggest that the overall shape of the synthetic spectra are in fact reliable, at least for the mass ($\sim$ 0.088 $M_{\odot}$) and metallicity ([Fe/H] = --0.71) of HD 114762B.  

Additionally, we calculated the bolometric luminosities and bolometric corrections for four known ultracool subdwarfs (LHS 377, LSR 2036+5059, SSSPM 1013--1356, and 2MASS 1626+3945) with parallaxes and whose near-infrared spectra were available in the SpeX Prism Spectral Library.  Ultracool subdwarfs have higher luminosities and smaller bolometric corrections than do ultracool dwarfs in the same spectral subclass.  While unresolved binarity is a possibility, a hotter effective temperature scale and/or a larger radius is a more likely explanation for the overluminosity.

Finally, we have also developed a  technique to estimate the distances to ultracool subdwarfs based on the available models and a single near-infrared spectrum.  Using the effective temperature from the best-fitting synthetic spectrum and an assumed age ($\sim$ 10 Gyr for objects with thick disk or halo kinematics), a radius can be inferred from evolutionary models.  The radius can then  be used along with the model scaling factor $\overline{C}_k$ (equal to $R^2/d^2$) to derive a distance estimate to the object.  We applied this technique to five ultracool subdwarfs with trigonometric distances and obtained distance estimates accurate to $<$ 10\% of the parallactic distances, or about 2 times better than estimates based on photometry alone.  This technique is particularly useful for ultracool subdwarfs, whose ages can be constrained by their kinematics.

It will be worthwhile to study HD 114762B at other wavelengths, including obtaining thermal infrared photometry ($\lambda$ $>$ 3 $\mu$m) to better constrain the long-wavelength end of the model-fitting, and, if possible, optical spectra, which would allow for an independent spectral classification of this object.  The dominant error in our luminosity determination originates from the errors in the photometry used to flux calibrate the spectra.  Better near-infrared photometry would therefore enable more a precise luminosity estimate and, by extension, more precise predictions from the evolutionary models.

 \acknowledgments
We thank the anonymous referee for helpful feedback.  We also thank Adam Burgasser for maintaining the SpeX Prism Spectral Library, as well as Peter Hauschildt and the entire PHOENIX group for making their models publicly available; likewise we thank Isabelle Baraffe, Gilles Chabrier, France Allard, and Peter Hauschildt for the public release of their evolutionary models.  This research has made use of the SIMBAD database, operated at CDS, Strasbourg, France.   This publication also makes use of data products from the Two Micron All Sky Survey, which is a joint project of the University of Massachusetts and the Infrared Processing and Analysis Center/California Institute of Technology, funded by the National Aeronautics and Space Administration and the National Science Foundation.  MCL and BPB acknowledge financial support from the Alfred P. Sloan Research Fellowship, NSF grant AST-0507833.  The authors wish to recognize and acknowledge the very significant cultural role and reverence that the summit of Mauna Kea has always had within the indigenous Hawaiian community.  We are most fortunate to have the opportunity to conduct observations from this mountain.

\facility{{\it Facilities:} \facility{Keck: II (OSIRIS)}, \facility{IRTF (SpeX)}}

\newpage

 \bibliographystyle{apj}
 \bibliography{sept09}

\end{document}